\documentclass[1p]{elsarticle}

\usepackage{lineno}
\modulolinenumbers[5]

\usepackage{amsmath,amssymb,amsfonts,mathtools}
\usepackage{algorithmic}
\usepackage[font=small,labelfont=bf]{caption}
\usepackage{subcaption}
\usepackage{graphicx}
\usepackage{dsfont}
\usepackage{tikz}
\usepackage{pgfplots}
\pgfplotsset{compat=1.9}
\usepackage[english]{babel}
\usepackage{multirow}
\usepackage{textcomp}
\usepackage{xcolor}

\usepackage{booktabs}
\usepackage{soul}
    
\biboptions{sort&compress}

\DeclareMathOperator{\sigmoid}{sig}

\newcommand{\vect}[1]{\mathbf{#1}}%
\newcommand{\mat}[1]{#1}
\definecolor{darkgreen}{rgb}{0, 0.5, 0}
\newcommand{\eg}{\emph{e.g.,}}

\journal{Journal of Optical Fiber Technology}

\begin{document}

\begin{frontmatter}

\centerline {(\emph{Invited Paper})}

\title{Complexity Reduction over Bi-RNN-Based Nonlinearity Mitigation in Dual-Pol Fiber-Optic Communications via a CRNN-Based Approach\tnoteref{t1}}

\author{Abtin Shahkarami}
\ead{abtin.shahkarami@telecom-paris.fr}
\author{Mansoor  Yousefi}
\ead{yousefi@telecom-paris.fr}
\author{Yves Jaouen \corref{cor1}}
\ead{yves.jaouen@telecom-paris.fr}

\address{Telecom Paris, Institut Polytechnique de Paris, 19 Place Marguerite Perey 91120 Palaiseau,  France}

\cortext[cor1]{Corresponding author}

\begin{abstract}
Bidirectional recurrent neural networks (bi-RNNs), in particular bidirectional long short term memory (bi-LSTM), bidirectional gated recurrent unit, and convolutional bi-LSTM models, have recently attracted attention for nonlinearity mitigation in fiber-optic communication. The recently adopted approaches based on these models, however, incur a high computational complexity which may impede their real-time functioning. In this paper, by addressing the sources of complexity in these methods, we propose a more efficient network architecture, where a convolutional neural network encoder and a unidirectional many-to-one vanilla RNN operate in tandem, each best capturing one set of channel impairments while compensating for the shortcomings of the other. We deploy this model in two different receiver configurations. In one, the neural network is placed after a linear equalization chain and is merely responsible for nonlinearity mitigation; in the other, the neural network is directly placed after the chromatic dispersion compensation and is responsible for joint nonlinearity and polarization mode dispersion compensation.
For a 16-QAM 64 GBd dual-polarization optical transmission over $14\times 80~{\rm km}$ standard single-mode fiber, we demonstrate that the proposed hybrid model achieves the bit error probability of the state-of-the-art bi-RNN-based methods with greater than $50\%$ lower complexity, in both  receiver configurations.
\end{abstract}


\end{frontmatter}


\section{Introduction}

The achievable spectral efficiency in fiber-optic communications is hampered by various channel impairments, such as chromatic dispersion (CD), polarization-mode dispersion (PMD), and Kerr nonlinearity \cite{kikuchi2015fundamentals,essiambre2010capacity,agrawal2012fiber}. CD and PMD can be compensated via linear equalization in a low-complexity regime, but nonlinear effects require nonlinear equalizers to be compensated \cite{cartledge2017digital,winzer2018fiber,savory2010digital,NFT}. Digital backpropagation (DBP) \cite{4738549,ssfm}, Volterra series transfer function \cite{guiomar2012mitigation,wang2015enhanced}, and optical phase conjugation \cite{fisher2012optical, jansen2006long, he2002optical} are among the conventional nonlinear equalizers in this area. These algorithms, however, require the knowledge of channel parameters and can be computationally complex, especially DBP.

In recent years, data-driven solutions, prompted by deep learning, have shown the potential to be satisfactory substitutes for the conventional algorithms \cite{8527529,8660506,shahkarami2021efficient,shahkarami2021attention,freire2021experimental,survey-yves,eriksson2017applying,koike2020neural}. These methods aim to achieve the performance of the conventional solutions with lower complexity and better generalizability over channel parameters. These solutions leverage neural networks made up of concatenated linear and nonlinear operators that can be optimized via the backpropagation learning algorithm.

A number of neural network structures have been investigated in the literature in this regard \cite{MLP-Yves,8386096,liu2021bi,deligiannidis2021performance,pedro2021perf&comp,Hager2018conf,sidelnikov2021advanced,wang2020bilstm,xu2021joint}. As universal approximators, multilayer perceptrons (MLPs) are among the first adopted models in this field. A number of MLP-based adopted models for equalization are discussed in \cite{MLP-Yves,ann-mlp-icocn,sidelnikov2018equalization,jiang2021solving}.
However, as MLPs try to find the correlation among each pair of the samples in data repeatedly in each layer using fully-connected layers, they are prone to over-fitting due to the large number of trainable parameters. Subsequently, they also bring about a relatively large number of floating-point operations (FLOPs).
Motivated by this problem, the hybrid of the convolutional neural network (CNN) and MLP (CNN+MLP) -based models have been studied \cite{8385711, 8386096, 8386099}. These models try to capture short-temporal dependencies among neighbouring symbols using convolution layers. Then, following these layers, they leverage a number of fully-connected layers to capture long-term dependencies. Due to the same reason as in MLPs, the fully-connected layers in CNN+MLP models are generally inefficient. On the other hand, while fully-convolutional models are able to capture long-term dependencies, they require large sequential kernels for this task, leading to high computational complexity. Over this, 
the authors of \cite{Hager2018conf, hager2018deep} propose a fully-convolutional model obtained via deep unfolding \cite{balatsoukas2019deep} of split-step Fourier method (SSFM); an approach termed Learned digital back-propagation (LDBP). LDBP is extended to dual-polarization (DP) in \cite{hager2020model} by integrating the distributed compensation of polarization-mode dispersion (PMD) into LDBP.
Yet, although LDBP achieves a favorable bit error rate (BER), it still suffers from a high computational complexity, which is rooted in the required number of spatial segments and the inefficiency of convolutional kernels to capture long-term features. 

Considering these problems,  bidirectional recurrent neural network (bi-RNN) -based approaches have recently piqued attention \cite{shahkarami2021attention}, \cite{deligiannidis2021performance,liu2021bi,pedro2021perf&comp}, \cite{deligiannidis2020compensation,gagne2021recurrent, karanov2019end,karanov2020experimental}. Bi-RNNs are efficient in capturing long-term dependencies owing to their ability to keep track of effective information over large sequences. 
This enables them to span a wide channel memory for equalization. Bi-RNNs, however, still require high computing resources \cite{pedro2021perf&comp,freire2021experimental}, which may hinder their implantation in real-time systems.

\begin{figure}[t]
    \centering
    \includegraphics[width=0.98\textwidth]{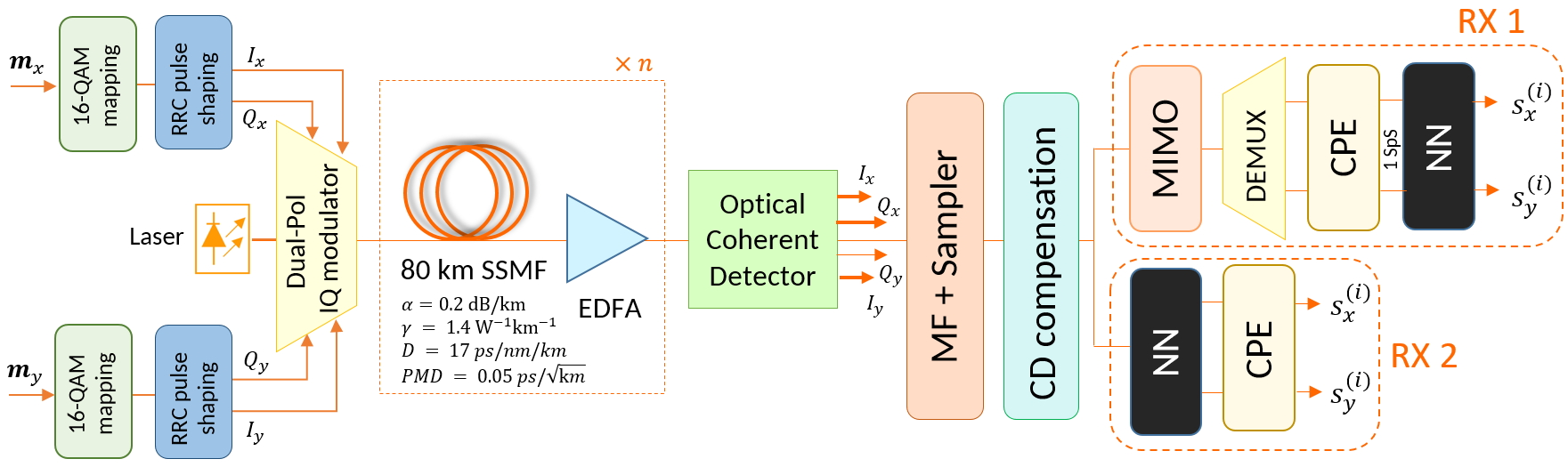}
    \caption{Block diagram of the considered fiber-optic transmission system. The top and bottom branches of the receiver represent the RX model 1 and 2, respectively. 
    }
    \label{fig: system-model}
\end{figure}

We note three issues concerning the computational complexity of these neural network equalizers. First, a bi-RNN is not an optimal tool for capturing short-term features. Second, in the gated RNNs, particularly long-short term memory (LSTM) and gated recurrent unit (GRU) structures, although the supplementary gates strengthen the neural network in the training mode, their complexity and overhead stay with the model in the inference mode as well. Third, having one RNN cell per time-step in the signal, as in \cite{pedro2021perf&comp, freire2021experimental, deligiannidis2020compensation, deligiannidis2021performance}, results in a high number of RNN cells, which brings about a hefty computational load. This complexity is doubled furthermore by the bidirectional processing flow. Besides, in the case of joint nonlinearity-PMD compensation, the computational burden on the recurrent layer grows in proportion to the signal's dimensionality.

To address these concerns, in this paper, we propose a hybrid architecture comprising a CNN encoder and a unidirectional many-to-one simple (vanilla) RNN working in tandem. The CNN component captures short-term dependencies using convolutional kernels. Furthermore, in parallel, by taking the data into a latent space using a balanced multi-channeling and striding,  the dimensionality of the signal is reduced towards 1 sample/symbol (SpS). The model groups the neighbour samples whose dependencies have been captured into distinct blocks to be passed as the time-steps to the RNN. The RNN, then, detects the long-range dependencies with low complexity. As the number of time-steps is substantially shrunk and the short-term features have been extracted, a unidirectional many-to-one vanilla RNN is sufficient for this task.

We apply the proposed model in two receiver (RX) configurations corresponding to the top and bottom branches of the block diagram in Fig. \ref{fig: system-model}. In the top processing branch (RX 1), the neural network is placed after the linear equalization processing chain to mitigate dual-polarization nonlinearity. In the bottom processing branch (RX 2), the neural network is placed directly after the CD compensation block to jointly mitigate PMD and nonlinearity. We demonstrate that for a 16-QAM 64 GBd dual-polarization optical transmission over $14 \times 80$ km standard single-mode fiber (SSMF), the proposed convolutional recurrent neural network (CRNN) model achieves the same performance as the bi-RNN, bi-GRU, bi-LSTM, and CNN+bi-LSTM models adopted in \cite{deligiannidis2021performance, liu2021bi, deligiannidis2020compensation, pedro2021perf&comp}, with at least $50\%$ fewer number of FLOPs.   

We highlight that in this study, in supplement to transmission system models considered in the mentioned bi-RNN-based studies, we also take into account the distributed PMD along the fiber link. 
Furthermore, we introduce lasers at the transmitter (TX) and RX, and study the impact of the laser phase noise on the performance of the neural networks.

This paper is organized as follows. Section \ref{sec: principles} discusses the fiber-optic transmission system model, the two considered receiver configurations, and the problem formulation. Section \ref{sec: recurrent-networks} provides a brief overview of the RNN, LSTM, and GRU structures and discusses their complexities measured by the number of FLOPs. The proposed hybrid neural network model is described in Section \ref{sec: solution}. Section \ref{sec: complexity} presents the results of the numerical simulations on comparing BER and complexity of the equalizers, and Section \ref{sec: conclusions} concludes the paper.


\section{Polarization-multiplexed fiber-optic transmission system}

\label{sec: principles}

This section describes the dual-polarization optical fiber communication system considered in this study, illustrated in Fig. \ref{fig: system-model}. In what follows, we describe the transmitter, optical fiber channel,  amplification model, and receiver. 
The description provided in this section focuses on the $x$ polarization; the equations for the $y$ polarization are similar upon swapping $x$ and $y$.

\subsection{Transmitter model} 

At the transmitter,
a bit stream $\vect{m_x} = (m^{(1)}_x,m^{(2)}_x,\dots,m^{(N_b)}_{x})$,  $m_x^{(i)} \in \{0,1\}$, is mapped to a sequences of symbols $\vect{s_x}=(s^{(1)}_x,s^{(2)}_x,..,s^{(N_s)}_{x})$, where $s_x^{(i)}$ are drawn from a 16-QAM constellation. The sequence of symbols $\vect{s_x}$ is then modulated to a digital waveform 
$q_x(t, 0) = \sum_{i=1}^{N_s} s_x^{(i)} p(t-i/R_s)$, where $p(t)$ is the root-raised-cosine (RRC) pulse shape and $R_s$ is the baud rate. 
This is similarly done for the bit stream $\vect{m_y}$ which is modulated to $q_y(t,0)$.  The waveforms
$q_x(t, 0)$ and $q_y(t, 0)$ are then multiplexed into an electric field 
with a dual-polarization Mach-Zehnder in-phase (I) and quadrature (Q) modulator, that is transmitted over  optical fiber. The modulator is driven by a laser with linewidth $\Delta\nu$ that introduces phase noise modeled by a Wiener process and Lorentzian power spectral density \cite[Chap.~3.5]{agrawal2012fiber}. The same realization of the phase noise is applied to  both polarizations.

\subsection{Dual-polarization channel model}

We consider an optical fiber link consisting of $N_{sp}$ spans of SSMF,  with  Erbium-doped fiber amplifiers (EDFAs) at the end of each span to compensate for the attenuation.

Propagation of signals in two polarizations of the electric field over one span of SSMF is modeled by the coupled nonlinear Schr\"odinger equation (CNLSE) \cite{CNLSE1,CNLSE2},  which describes 
the interactions between the two states of polarization along the fiber.
The equation for the $x$-polarization is \cite[Chap.~6.1]{agrawal2012fiber}
\begin{equation}
 \frac{\partial  q_x(t,z)}{\partial z} = -\frac{\alpha}{2} q_x -\beta_{1x}\frac{\partial  q_x}{\partial t}-\frac{j\beta_2}{2}\frac{\partial^2  q_x}{\partial t^2}
     + j\gamma \Bigl(| q_x|^2+\frac{2}{3}| q_{y}|^2\Bigr)  q_x.
\label{eq: cnlse}
\end{equation}
Here, $q_x(t,z)$ is the complex envelope of the signal in the $x$ polarization propagating in the fiber as a function of time $t$ and distance $z$, $\alpha$ and $\beta_2$ are respectively the attenuation and CD coefficients, and $\gamma$ is the nonlinearity parameter. The first-order dispersion coefficient $\beta_{1x}$ depends on the polarization, giving rise to the differential group delay (DGD) and PMD.
\cite[Sec. VI.C]{kikuchi2015fundamentals}. The equation for the $y$-polarization, as mentioned, is similar to \eqref{eq: cnlse} upon swapping $x$ and $y$ in \eqref{eq: cnlse}.

Equation \eqref{eq: cnlse} is numerically solved using the split-step Fourier method (SSFM) \cite{ssfm} with distributed PMD as follows. The fiber span is divided into $K$ segments of length $\delta$. 
In each segment $i$, a linear, PMD, and nonlinear step is performed consecutively as follows.

\paragraph*{\textnormal{1.} Linear step} Loss and CD are applied in the frequency domain to the signal as
\begin{equation}
\hat{q}_x(\omega, z) \mapsto 
  \exp
    \left (
    -\frac{\alpha}{2}\delta + j\frac{\beta_2}{2}{\omega}^2 \delta
    \right )\hat{q}_x(\omega, z),
\end{equation}
where $\hat{q}_x(\omega, z)$ is the Fourier transform of $q_x(t,z)$.

\paragraph*{\textnormal{2.} PMD step}

To model the distributed PMD, a unitary matrix $\mat{J}^{(i)}(\omega)$ is applied to the signal vector $\vect{q}(t,z) = [q_x(t,z),q_y(t,z)]^{\textnormal{T}}$ in the frequency domain as
\begin{equation}
\hat{\vect{q}}(\omega, z) \mapsto 
     \mat{J}^{(i)}(\omega)\hat{\vect{q}}(\omega, z),
\end{equation}
where $\hat{\vect q}(\omega, z)$ is the Fourier transform of $\vect q(t,z)$, and
\begin{equation}
     \mat{J}^{(i)}(\omega) =
     \mat{R}^{(i)}\mat{D}^{(i)}(\omega),
\end{equation}
in which 
\begin{equation}
     \mat{R}^{(i)}  =
     \begingroup
     \begin{pmatrix} e^{j\frac{\phi_i}{2}}\cos(\theta_i) & e^{-j\frac{\phi_i}{2}}\sin(\theta_i)\\
     -e^{j\frac{\phi_i}{2}}\sin(\theta_i) & e^{-j\frac{\phi_i}{2}}\cos(\theta_i) 
     \end{pmatrix},
     \endgroup
\end{equation}
is the rotation matrix, where $(\theta_i)_{i=1}^K$ and $(\phi_i)_{i=1}^K$ are sequences of independent and identically distributed (iid) 
random variables drawn from a uniform distribution on $[0,2\pi)$. 
In consequence, the state of polarization (SOP) is uniformly distributed over the surface of the Poincar\'e spher along the fiber.
Further, $\mat{D}^{(i)}(\omega)$ is the DGD operator
\begin{equation}
    \mat{D}^{(i)}(\omega) = \begin{pmatrix} e^{-j\omega\frac{\tau^{(i)}}{2}} & 0
    \\ 0 & e^{j\omega\frac{\tau^{(i)}}{2}}\end{pmatrix},
\end{equation}
where $ (\tau^{(i)})_{i=1}^K$ are DGD parameters, which we assume to be a sequence of iid random variables drawn from the
probability distribution $N(0,\tau \sqrt{\delta})$, where $\tau$ (measured in ${\rm ps/\sqrt{km}}$) is the PMD parameter. 
\par

\paragraph*{\textnormal{3.} Nonlinear step} Kerr nonlinearity effect for $x$-polarization is modeled in the time domain as
\begin{equation}
q_x(t,z)\mapsto 
    \exp
    \left (
    j\gamma \delta\Bigl(| q_x(t,z)|^2+\frac{2}{3}|  q_{y}(t,z)|^2 \Bigr)
    \right ) q_x(t,z).
\label{eq: nonlinearity}    
\end{equation}
For the $y$-polarization, the process is similar by swapping $x$ and $y$ in \eqref{eq: nonlinearity}. 
\par

\subsection{Receiver models} 
\label{sec: rx-models-dscrptn}

At the receiver, an optical coherent detector translates the optical signal to four electrical signals, corresponding to the I and Q components of each polarization. It is assumed that the lasers at TX and RX operate at the same frequency, i.e., the carrier frequency offset (CFO) is zero. Consecutively, the sampled waveforms are forwarded to the digital signal processing (DSP) chain to compensate for the channel impairments.

The first step in the DSP is CD compensation, reversing the dispersion effect throughout the fiber in one-shot  as
\begin{equation*}
\hat{q}_x(\omega, \mathfrak L) \mapsto 
    \exp \left (
     -j\frac{\beta_2}{2}\omega^2 \mathfrak{L}
    \right ) \hat{q}_x(\omega, \mathfrak{L}),
\end{equation*}
where $\mathfrak{L}$ is the fiber length. Next, depending on the task of the neural network, two receiver configurations, labeled as RX 1 and RX 2 in Fig. ~\ref{fig: system-model}, are considered.

\paragraph{RX model 1}

The neural network is placed after the linear equalization, with the aim of mitigating nonlinear channel impairments. 
The linear DSP  consists of a cascade of CD compensation, a radius-directed-equalization (RDE) \cite{fatadin2009blind} -based multiple-input-multiple-output (MIMO) algorithm to compensate for the PMD, a demultiplexer to separate the two polarizations, and a two-stage carrier phase estimation (CPE) algorithm to compensate for the phase offset \cite{pfau2009PN}. 
The CFO compensation is not required, since the CFO is assumed to be zero.

\paragraph{RX model 2}

The neural network is placed after the CD compensation, and its purpose is to jointly mitigate the nonlinearity and PMD. There is a CPE block after the neural network in this RX architecture to compensate for the laser phase noise, as the neural network is not able to compensate for the phase noise efficiently due to its randomness.

\begin{figure}[!ht]
    \centering
    \includegraphics[width=0.6\textwidth]{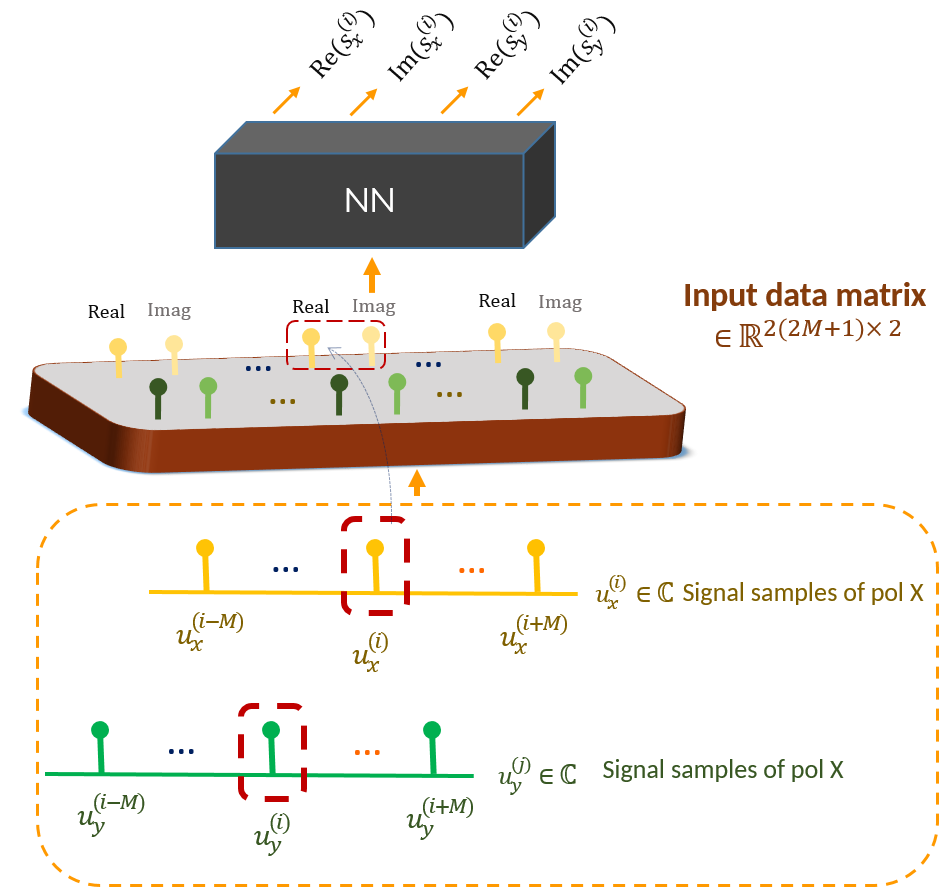}
    \caption{Input-output model of the neural networks in both RX models.}
    \label{fig: input-output-scheme}
\end{figure}

The input-output model of the neural networks in both RX models is shown in Fig.~\ref{fig: input-output-scheme}. To equalize the symbols $s_x^{(i)}$ and $s_y^{(i)}$
at time-step $i$, the neural network processes two vectors $\vect u_x^{(i)}$  and $\vect u_y^{(i)}$ containing a window of the time-domain signal samples from each polarization output by the previous block in the DSP chain, \eg\
\begin{equation*}
\vect u_x^{(i)}= \Bigl(u_x^{(i- M)},\cdots, u_x^{(i)},\cdots, u_x^{(i+ M)}\Bigr)^T.
\end{equation*}
The input vectors are centered at time-step $i$, and span $2M$ neighbor samples ($M$ left, $M$ right).
If the effective channel memory in number of symbols is $\bar M$, then
$ M= \bar M \textnormal{SpS} + (\textnormal{SpS} - 1)/2 $.
The real and imaginary parts of the samples in $\vect u_x^{(i)}$ and $\vect u_y^{(i)}$ are then split and placed in every other position in a corresponding vector, \eg
\begin{equation*}
\tilde{\vect u}_x^{(i)}= \Bigl(
\Re(u_x^{(i- M)}), 
\cdots, 
\Re(u_x^{(i)}),\Im(u_x^{(i)}),
\cdots, \Im(u_x^{(i+ M)})\Bigr)^T.
\end{equation*}
Consequently, the vectors $\tilde{\vect u}_x^{(i)}$ and $\tilde{\vect u}_y^{(i)}$ are stacked together,  and the matrix $\mat{U}^{(i)}=(\tilde{\vect u}_x^{(i)}, \tilde{\vect u}_y^{(i)}) \in \mathbb{R}^{2(2M + 1)\times 2}$ is passed to the neural network as input.
For RX 1, the SpS at the input of the neural network is  1. 
After processing, the neural network outputs the real and imaginary parts of the equalized symbol in the $x$- and $y$- polarization at time-step $i$ (4 outputs). 

The neural networks are trained with $\mat{U}^{(i)}$ as the input data matrix, and the corresponding correct transmitted symbols $s_x^{(i)}$ and $s_y^{(i)}$ in the standard 16-QAM constellation as the ground truth.
The performance of the receivers is measured in the test mode by the Q-factor
\begin{equation}
    \textnormal{Q-factor} = 20 \log_{10}[\sqrt{2}\textnormal{erfc}^{-1}(2\textnormal{BER})],
\end{equation}
where BER is the bit-error-rate, and erfc is the complementary error function.

\section{Learning by recurrent neural networks}

\label{sec: recurrent-networks}

RNNs are a category of neural networks designed to operate on temporal sequences of data with correlated samples. These neural networks are composed of recurrent cells whose states are influenced not only by the current input but also by the past or even future (for bi-RNNs) time-steps. This scheme is enabled by the so-called memory of the RNN, which is emulated by the hidden state property. That is to say, the sequential information is maintained in the RNN’s hidden state, which is utilized to positively affect the processing of each new time-step as the sequence steps forward.

A unidirectional many-to-many vanilla 
RNN layer is made up of a number of recurrent cells, governed by the equations
\begin{align}
    \label{eq: RNN-formula}
    \vect{h}^{(t)} &= \sigma_1(\mat{W_h} \vect{h}^{(t-1)} + \mat{W_x} \vect{x}^{(t)} + \vect{b_h} ), 
 \\
 \vect y^{(t)} &= \sigma_2(\mat{W_y} \vect{h}^{(t)} + \vect {b_y}), \nonumber 
\end{align}
where $\vect{x}^{(t)} \in\mathbb{R}^{n_i}$ and $\vect{h}^{(t)}\in\mathbb{R}^{n_h}$ are respectively the input and the hidden state of time-step $t$; $\mat{W_h} \in \mathbb{R}^{n_h \times n_h}$, $\mat{W_x} \in \mathbb{R}^{n_h \times n_i}$, 
$\mat{W_y} \in \mathbb{R}^{n_y \times n_h}$  are the weight matrices, $\vect{b_h} \in \mathbb{R}^{n_h}$ and $\vect{b_y} \in \mathbb{R}^{n_y}$ are the bias vectors, and $\sigma_1(.)$ and $\sigma_2(.)$ are the activation functions. 

Thanks to the memory property of RNNs, enabled by their hidden state, they can capture long-term dependencies among the temporal sequence.
However, a major challenge with RNNs is the problem of vanishing gradients in the backpropagation, which limits the performance of RNNs for large numbers of time-steps. Several variants of RNNs have been proposed to mitigate this problem and enable the processing of longer sequences.

LSTM and GRU networks are primary variants in this respect \cite{yu2019review}. These networks incorporate gates in the RNN cells to regulate the flow of information.
These gates learn which information should be kept or discarded in a time-series, allowing the relevant information to be preserved throughout a long sequence of RNN cells. 

\subsection{Long-short term memory networks}

LSTM networks incorporate three supplementary gates in addition to a cell state to facilitate handling large sequences: update gate $\Gamma_u$, forget gate  $\Gamma_f$, and output gate $\Gamma_o$. The cell state acts as an additional supportive memory that keeps track of the relevant information throughout the process flow.

The forget gate determines what information should be discarded and what information must be retained from the previous cell state. It jointly processes the previous hidden state as well as the current input and passes the resulting feature through the sigmoid function resulting in the values in the interval (0,1). The closer this value is to 0, the more the corresponding information is forgotten, and vice versa. 
Similarly, the update and output gates learn, respectively, what values in the cell state should be updated and what information in the current cell state should be kept as the output of the hidden state. 
The equations describing an LSTM cell in the time-step $t$ are as follows:
\begin{align}
\label{eq: LSTM-formulas}
     & \vect{\Tilde{c}}^{(t)} = \sigma_{\tanh} \left( \mat{{W}_{ch}} \vect{h}^{(t-1)} + \mat{{W}_{cx}} \vect{x}^{(t)} + \vect{b_c} \right) \nonumber \\
     & \vect {\Gamma_u} =  \sigma_{\sigmoid} \left( \mat{W_{uh}} \vect{h}^{(t-1)} + \mat{{W}_{ux}} \vect{x}^{(t)} + \vect{b_u} \right) \nonumber \\
     & \vect {\Gamma_f} =  \sigma_{\sigmoid} \left( \mat{W_{fh}} \vect{h}^{(t-1)} + \mat{{W}_{fx}} \vect{x}^{(t)} + \vect{b_f} \right) \\
     & \vect {\Gamma_o} =  \sigma_{\sigmoid} \left( \mat{W_{oh}} \vect{h}^{(t-1)} + \mat{{W}_{ox}} \vect{x}^{(t)} + \vect{b_o} \right) \nonumber \\
     & \vect{c}^{(t)} = \vect {\Gamma_u} \odot \vect{\Tilde{c}}^{(t)} + \vect {\Gamma_f} \odot \vect{c}^{(t-1)}  \nonumber \\
     & \vect{h}^{(t)} = \vect {\Gamma_o} \odot \sigma_{\tanh}(\vect{c}^{(t)}) \nonumber
\end{align}
where $\vect{x}^{(t)}\in\mathbb{R}^{n_i}$ is input, $\vect{c}^{(t)}\in\mathbb{R}^{n_c}$ is cell state, $\vect{h}^{(t)}\in\mathbb{R}^{n_h}$ is hidden state, $\mat{{W}_{ch}}, \mat{W_{uh}}, \mat{W_{fh}}, \mat{W_{oh}} \in \mathbb{R}^{n_h \times n_h}$ and $\mat{{W}_{cx}}, \mat{{W}_{ux}}, \mat{{W}_{fx}}, \mat{{W}_{ox}} \in \mathbb{R}^{n_h \times n_i}$ are weight matrices, 
$\vect{b_c}, \vect{b_u}, \vect{b_f}, \vect{b_o} \in \mathbb{R}^{n_h}$ are biases,
and $\sigma_{sig}$ and $\sigma_{tanh}$ are the sigmoid and tanh activations, respectively. $\odot$ is the Hadamard product.

\subsection{Gated recurrent unit networks}

The GRU networks serve the same purpose as LSTM networks. In GRU units, there exist two gates, compared to three in LSTM, and the cell state equals the hidden state. Relevance (or reset) gate $\Gamma_r$ and update $\Gamma_u$ gate are the two gates working almost similar to forget and update gates in LSTM. The mathematical formulation of a GRU unit is as follows:
\begin{align}
\label{eq: GRU-formulas}
     &\vect{\Tilde{c}}^{(t)} = \sigma_{tanh} \left( \mat{{W}_{cc}}(\vect {\Gamma_r} \odot \vect{c}^{(t-1)}) + \mat{{W}_{cx}} \vect{x}^{(t)} + \vect{b_c} \right) \nonumber \\
    & \vect \Gamma_u =  \sigma_{sig} \left( \mat{W_{uc}} \vect{c}^{(t-1)} + \mat{{W}_{ux}} \vect{x}^{(t)} + \vect{b_u} \right) \nonumber \\
    & \vect \Gamma_r =  \sigma_{sig} \left( \mat{W_{rc}} \vect{c}^{(t-1)} + \mat{{W}_{rx}} \vect{x}^{(t)} + \vect {b_r} \right)  \\
    & \vect{c}^{(t)} = \vect {\Gamma_u} \odot \vect{\Tilde{c}}^{(t)} + (1 - \vect {\Gamma_u}) \odot \vect{c}^{(t-1)}  \nonumber \\
    & \vect{h}^{(t)} = \vect{c}^{(t)}, \nonumber
\end{align}
where $\vect{x}^{(t)}$, $\vect{c}^{(t)}$ and $\vect{h}^{(t)}$ have the same definition as in LSTM; 
$\mat{W_{cc}}, \mat{W_{uc}}, \mat{W_{rc}} \in \mathbb{R}^{n_h \times n_h}$, $\mat{{W}_{cx}}, \mat{{W}_{ux}}, \mat{{W}_{rx}} \in \mathbb{R}^{n_h \times n_i}$, and $\vect{b_c}, \vect{b_u}, \vect {b_r} \in \mathbb{R}^{n_h}$.

GRUs have been shown to be able to achieve comparable performance as LSTM in several applications, such as speech recognition, traffic load prediction, etc., with lower computational complexity \cite{ravanelli2018light,luo2021uav,nadig2019april}.

\subsection{Complexity analysis}
\label{sec:complexity}

In this section, we analyze the computational complexity of LSTM, GRU, and vanilla RNN networks.

We measure the computational complexity of a neural network in the inference phase by the number of floating-point real additions and multiplications (FLOPs).
From \eqref{eq: RNN-formula}, the number of FLOPs performed in a vanilla RNN cell in one time-step is
\begin{equation}
    \textnormal{FLOPs}^{\{\textnormal{RNN}\}} = n_i n_h + n_{h}^2 +2n_h + \eta n_h,
\label{eq: FLOPS-RNN}
\end{equation}
where $\eta$ is the number of FLOPs that the activation takes to process the input value.

Similarly, by \eqref{eq: LSTM-formulas} and \eqref{eq: GRU-formulas}, the number of FLOPs in an LSTM and GRU cell are
\begin{equation}
    \textnormal{FLOPs}^{\{\textnormal{LSTM}\}} = 4 \times [n_i n_h + n_{h}^2 +3n_h] 
    + 5\eta n_h,
\label{eq: FLOPS-LSTM}
\end{equation}
and
\begin{equation}
    \textnormal{FLOPs}^{\{\textnormal{GRU}\}} = 3 \times [n_i n_h + n_{h}^2 +2n_h + \eta n_h] + 5n_h,
\label{eq: FLOPS-GRU}
\end{equation}
respectively.
The last $n_h$ in \eqref{eq: FLOPS-GRU} is due to the $\vect \Gamma_r \odot \vect{c}^{(t-1)}$ operation in the calculation of $\vect{\Tilde{c}}^{(t)}$. In \eqref{eq: FLOPS-RNN}-\eqref{eq: FLOPS-GRU}, it is assumed that tanh and sigmoid activations incur the same number of FLOPs.

The number of FLOPs of a recurrent layer equals the number of FLOPs in one time-step multiplied by the number of time-steps. 
In addition, for bidirectional layers, this number should be multiplied by a factor of 2.

\section{Reducing complexity using latent space of CNNs}
\label{sec: solution}

\begin{figure}
    \centering
    \includegraphics[width=0.8\textwidth]{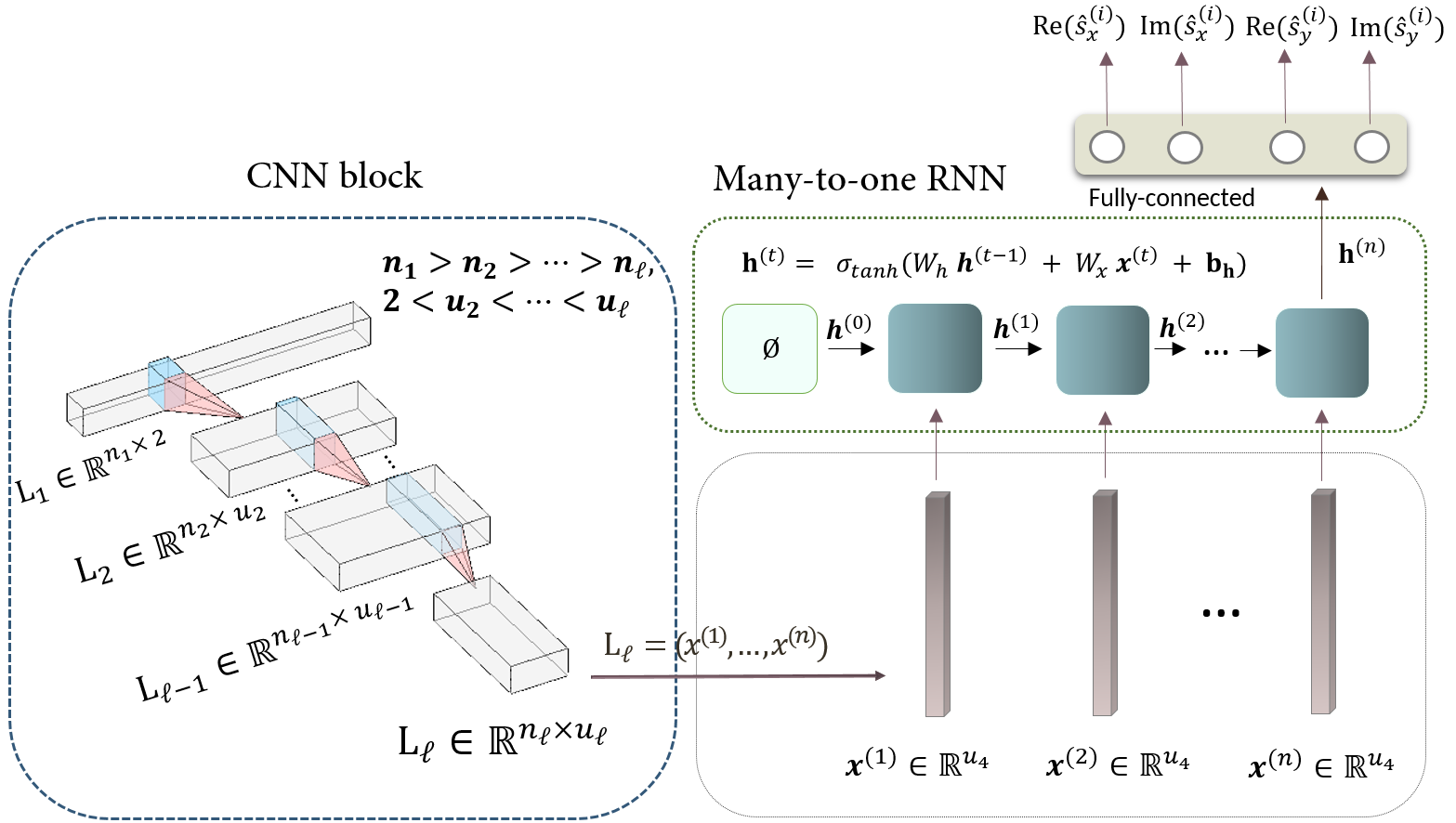}
    \caption{Schematic illustration of the process flow in the proposed CRNN model. 
    }
    \label{fig: cnn+rnn+flow}
\end{figure}

Depending on distance and bandwidth, there are often long- and short-term dependencies in the received signal that need to be extracted for equalization. We note that although bidirectional recurrent networks are appropriate tools to capture long-distance features, using them to learn short-term dependencies brings about a high computational inefficiency. This inefficiency is mainly rooted in the high number of time-steps, the presence of recurrent gates, and the bidirectional processing flow. Besides, in RX 2, the dimensionality of the signal passed to the neural network is usually 2 SpS or higher. This oversampling imposes an additional computational burden on the RNN layer, as it can be seen in \eqref{eq: FLOPS-RNN}-\eqref{eq: FLOPS-GRU}.

To address these problems, in this study, we aim to place a pre-processing block prior to the RNN to eliminate the mentioned sources of complexity. To this end, we leverage a CNN structure because of its twofold advantages. Firstly, CNNs are established to be efficient tools to capture short-term features, which minimizes the overlap between short-term and long-term dependencies extraction. Secondly, by the multi-channeling and striding properties, they can take the data into a latent space where not only the neighbour samples whose dependencies have been captured are grouped in the depth-axis of the feature map, but also using appropriate hyper-parameters, the dimensionality of data is reduced towards 1 SpS (the latter is only applicable to RX 2).

This enables the RNN to take the vectors in the depth-axis of the feature-map provided by the CNN encoder, as the time-steps. This highly limits the number of time-steps in the RNN, facilitating the replacement of gated recurrent cells with vanilla RNN cells. In addition, there is no need for a high number of hidden units in the RNN cells since there is no need for further processing to capture dependencies within the input vector as they have already been extracted. These together enable us to replace the bidirectional recurrent layer with a unidirectional many-to-one vanilla RNN layer. This is owing to the fact that the effective information (RNN memory) can be maintained along the layer while it is updated at each time-step in a cascade of quite limited number of RNN cells (3 to 6). 
Note that in the absence of a CNN block in the model, besides the burden of capturing both types of dependencies on the uni-directional RNN layer, the RNN memory should have been maintained throughout the long cascade of RNN cells, making the neural network inefficient for real-time operation.

\begin{figure}[t]
    \centering
    \includegraphics[width=0.4\textwidth]{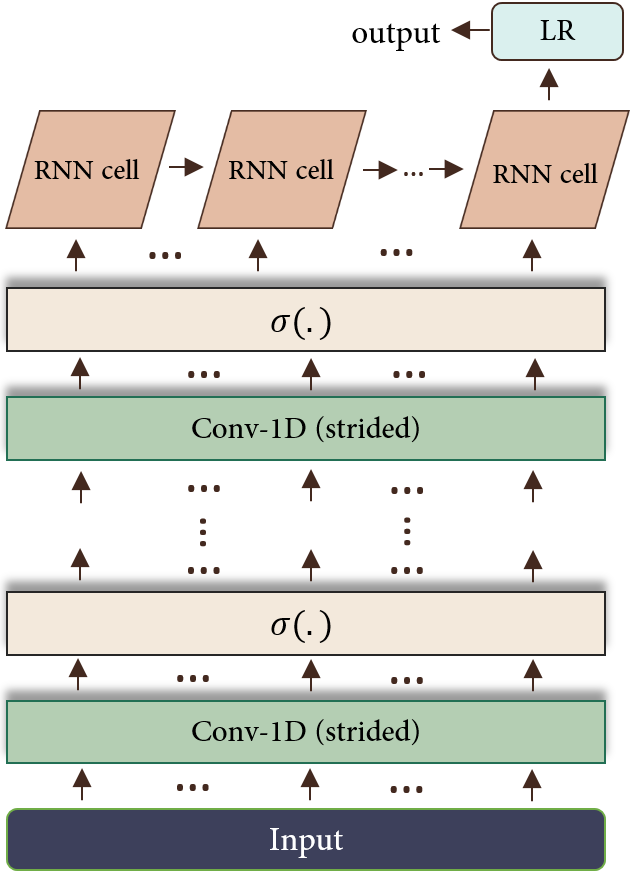}
    \caption{The architecture of the proposed CRNN model. The number of convolution layers is a hyper-parameter that is optimized.
    }
    \label{fig: cnn+rnn+block}
\end{figure}

Fig. \ref{fig: cnn+rnn+flow} illustrates the schematic of the proposed neural network model. As discussed, the neural network receives a data matrix in $\mathbb{R}^{2(2M + 1) \times 2}$ according to the input-output scheme discussed in Section. \ref{sec: principles}. This matrix is initially processed by a CNN structure. In the CNN, at each layer, the width of the data matrix (feature-map) is reduced (via striding), and the depth is increased (via applying filters). For RX 2, narrowing the width and increasing the depth should be in a  disproportionate manner to reduce the dimensionality (narrowing the width must be further). Following each layer, there is an activation function to introduce nonlinearity to the output. The feature-map output by the CNN is then passed to a unidirectional many-to-one vanilla RNN layer, where the vectors in the depth-axis are considered as the time-steps. According to the discussed logic, this recurrent layer is capable of capturing long-range dependencies while considering the short-term features. The final state of the recurrent layer (the hidden state output by the last RNN cell) is then linked to four processing units via a fully-connected layer, forming a linear regression model for each of the four units. These units are intended to output the real and imaginary parts of the symbols in the target position $i$, at x- and y- polarizations. The block diagram of the proposed model is also depicted in Fig. \ref{fig: cnn+rnn+block}.

Note that since the SOP frequently changes along the optical link in birefringent fibers, in RX 2 the neural network weights must be updated frequently to adapt the receiver to the channel. This is discussed more in detail in the next section.

\begin{table}[t]
\caption{Details of the hidden layers of the implemented neural networks for RX 1. The output layer for all the models is a linear fully-connected layer with 4 units.}
\centering
\footnotesize
\begin{tabular}{@{}cccc@{}}
\toprule
\textbf{Model}           & \textbf{Layer} & \textbf{Type} & \textbf{Details}                                                  \\ \midrule
\multirow{4}{*}{MLP \cite{MLP-Yves}}     & layer 1        & Dropout       & dropout rate: 0.4                                                 \\ \cmidrule(l){2-4} 
                         & layer 2,3      & FC            & \#units: 1536,  activ.: tanh                                      \\ \cmidrule(l){2-4} 
                         & layer 4        & Dropout       & dropout rate: 0.3                                                 \\ \cmidrule(l){2-4} 
                         & layer 5,6      & FC            & \#units: 1536,  activ.: tanh                                      \\ \midrule
\multirow{4}{*}{CNN+MLP \cite{8386096}}                                                     & layer 1 & Conv-1D & $\textnormal{L}_{ker}$:  49, strd: 1, \#ch.: 4, activ.: relu        \\ \cmidrule(l){2-4} 
                         & layer 2        & Conv-1D       & $\textnormal{L}_{ker}$:  49, strd: 1, \#ch.: 6, activ.: relu \\ \cmidrule(l){2-4} 
                         & layer 3        & Conv-1D       & $\textnormal{L}_{ker}$:  49, strd: 1, \#ch.: 8, activ.: relu \\ \cmidrule(l){2-4} 
                         & layer 4,5      & FC            & \#units: 768, activ.: tanh                                        \\ \midrule
\multirow{2}{*}{bi-LSTM \cite{deligiannidis2020compensation}} & layer 1        & bi-LSTM       & \#units: 144; activ.: tanh, sigmoid                               \\ \cmidrule(l){2-4} 
                         & layer 2        & Flattening    & data format: channel last                                         \\ \midrule
\multirow{2}{*}{bi-GRU \cite{liu2021bi}}  & layer 1        & bi-GRU        & \#units: 144; activ.: tanh, sigmoid                               \\ \cmidrule(l){2-4} 
                         & layer 2        & Flattening    & data format: channel last                                         \\ \midrule
\multirow{2}{*}{bi-RNN \cite{deligiannidis2021performance}}  & layer 1        & bi-RNN        & \#units: 240; activ.: tanh                                        \\ \cmidrule(l){2-4} 
                         & layer 2        & Flattening    & data format: channel last                                         \\ \midrule
\multirow{3}{*}{CNN+bi-LSTM \cite{pedro2021perf&comp}}                                                 & layer 1 & Conv-1D & $\textnormal{L}_{ker}$:  49, strd: 1, \#ch.: 2,  activ.: leaky relu \\ \cmidrule(l){2-4} 
                         & layer 2        & bi-LSTM       & \#units: 136,  activ.: tanh, sigmoid                              \\ \cmidrule(l){2-4} 
                         & layer 3        & Flattening    & data format: channel last                                         \\ \midrule
\multirow{4}{*}{\begin{tabular}[c]{@{}c@{}}\textbf{Proposed CRNN} \end{tabular}} & layer 1 & Conv-1D & $\textnormal{L}_{ker}$: 49, strd: 1, \#ch.: 4, activ: relu          \\ \cmidrule(l){2-4} 
                         & layer 2        & Conv-1D       & $\textnormal{L}_{ker}$: 9, strd: 9, \#ch.: 36, activ: relu        \\ \cmidrule(l){2-4} 
                         & layer 3        & Conv 1-D      & $\textnormal{L}_{ker}$:  5, strd: 5, \#ch.: 180, activ: relu      \\ \cmidrule(l){2-4} 
                         & layer 4        & uni-RNN       & \#units: 454, activ.: tanh                                        \\ \bottomrule
\end{tabular}
\label{tab: model-details-RX1}
\end{table}

\section{Performance and complexity comparison}
\label{sec: complexity}

We considered a DP-16-QAM 64 GBd point-to-point fiber-optic transmission system over $14 \times 80$ km SSMF optical-link, according to the system model illustrated in Fig. \ref{fig: system-model}, with the following parameters: fiber loss $a_{\rm{dB}} = 0.2$ dB/km, chromatic dispersion $D{=}17~{\rm ps/nm/km}$, nonlinearity parameter $\gamma = 1.4~{\rm W}^{-1}{\rm km}^{-1}$, PMD value $0.05~{\rm ps/\sqrt{km}}$, EDFA noise figure NF = 5 dB, and the laser linewidth $\Delta\nu=100$ kHz. For pulse shaping, RRC filters with a roll-off of 0.25 are employed. Forward propagation was simulated using SSFM with 8 SpS and 80 step/span (increasing either value did not affect the results). The sampling rate at RX was set to 2 SpS.

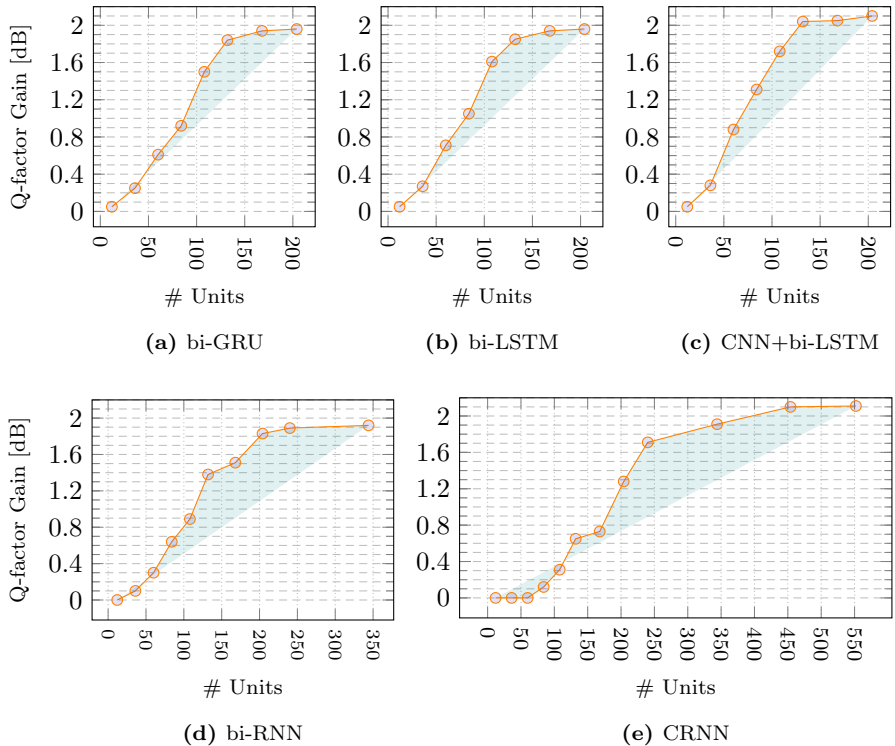
\begin{figure}[t]
\centering

\subfloat[bi-GRU]{%
\begin{tikzpicture}[trim axis left,trim axis right]
\begin{axis}[
        height=4.5cm, width=4.5cm,
        yminorgrids, tick align=inside,
        minor grid style={densely dashed},
        ymax=2.2,
        ytick={0,0.4,0.8,1.2,1.6,2.0},
        minor ytick={0,0.1,0.2,0.3,0.4,0.5,0.6,0.7,0.8,0.9,1.0,1.1,1.2,1.3,1.4,1.5,1.6,1.7,1.8,1.9,2.0,2.1,2.2},
        xtick={0,50,...,350},
                xmajorgrids, tick align=inside,
        major grid style={densely dotted},
        xticklabel style = {font=\footnotesize, rotate=-90},
                ylabel={\footnotesize Q-factor Gain [dB]},
        xlabel={\footnotesize \# Units},
]
\addplot+[draw=orange, fill=teal,  fill opacity=0.12] coordinates
	{(12,0.05)(36,0.25)(60,0.61)(84,0.92)(108,1.50)(132,1.84)(168,1.94)(204,1.96)};
\end{axis}
\end{tikzpicture}%
}
\hspace{0.75cm}
\subfloat[bi-LSTM]{%
\begin{tikzpicture}[trim axis left,trim axis right]
\begin{axis}[
        height=4.5cm, width=4.5cm,
        yminorgrids, tick align=inside,
        minor grid style={densely dashed},
        ymax=2.2,
        ytick={0,0.4,0.8,1.2,1.6,2.0},
        minor ytick={0,0.1,0.2,0.3,0.4,0.5,0.6,0.7,0.8,0.9,1.0,1.1,1.2,1.3,1.4,1.5,1.6,1.7,1.8,1.9,2.0,2.1,2.2},
        xtick={0,50,...,350},
                xmajorgrids, tick align=inside,
        major grid style={densely dotted},
        xticklabel style = {font=\footnotesize, rotate=-90},
        xlabel={\footnotesize \# Units},
]

\addplot+[draw=orange, fill=teal,  fill opacity=0.12] coordinates
	{(12,0.05)(36,0.27)(60,0.71)(84,1.05)(108,1.61)(132,1.85)(168,1.94)(204,1.96)};
	
\end{axis}
\end{tikzpicture}%
}
\hspace{0.75cm}
\subfloat[CNN+bi-LSTM]{%
\begin{tikzpicture}[trim axis left,trim axis right]
\begin{axis}[
        height=4.5cm, width=4.5cm,
        yminorgrids, tick align=inside,
        minor grid style={densely dashed},
        ymax=2.2,
        ytick={0,0.4,0.8,1.2,1.6,2.0},
        minor ytick={0,0.1,0.2,0.3,0.4,0.5,0.6,0.7,0.8,0.9,1.0,1.1,1.2,1.3,1.4,1.5,1.6,1.7,1.8,1.9,2.0,2.1,2.2},
        xtick={0,50,...,350},
                xmajorgrids, tick align=inside,
        major grid style={densely dotted},
        xticklabel style = {font=\footnotesize, rotate=-90},
        xlabel={\footnotesize \# Units},
]

\addplot+[draw=orange, fill=teal,  fill opacity=0.12] coordinates
	{(12,0.05)(36,0.28)(60,0.88)(84,1.31)(108,1.72)(132,2.04)(168,2.05)(204,2.10)};
	
\end{axis}
\end{tikzpicture}%
}

\vspace{0.5cm}

\subfloat[bi-RNN]{%
\begin{tikzpicture}[trim axis left,trim axis right]
\begin{axis}[
        height=4.5cm, width=5.55cm,
        yminorgrids, tick align=inside,
        minor grid style={densely dashed},
        ymax=2.2,
        ytick={0,0.4,0.8,1.2,1.6,2.0},
        minor ytick={0,0.1,0.2,0.3,0.4,0.5,0.6,0.7,0.8,0.9,1.0,1.1,1.2,1.3,1.4,1.5,1.6,1.7,1.8,1.9,2.0,2.1,2.2},
        xtick={0,50,...,350},
                xmajorgrids, tick align=inside,
        major grid style={densely dotted},
        xmajorgrids, tick align=inside,
        major grid style={densely dotted},
        xticklabel style = {font=\scriptsize, rotate=-90},
        ylabel={\footnotesize Q-factor Gain [dB]},
        ylabel near ticks,
        xlabel={\footnotesize \# Units},
]
\addplot+[draw=orange, fill=teal,  fill opacity=0.12] coordinates
	{(12,0.0)(36,0.1)(60,0.3)(84,0.64)(108,0.89)(132,1.38)(168,1.51)(204,1.83)(240,1.89)(344,1.92)};
\end{axis}
\end{tikzpicture}%
}
\hspace{0.75cm}
\subfloat[CRNN]{%
\begin{tikzpicture}[trim axis left,trim axis right]
\begin{axis}[
        height=4.5cm, width=7.27cm,
        yminorgrids, tick align=inside,
        minor grid style={densely dashed},
        ymax=2.2,
        ytick={0,0.4,0.8,1.2,1.6,2.0},
        minor ytick={0,0.1,0.2,0.3,0.4,0.5,0.6,0.7,0.8,0.9,1.0,1.1,1.2,1.3,1.4,1.5,1.6,1.7,1.8,1.9,2.0,2.1,2.2},
        xtick={0,50,...,550},
                xmajorgrids, tick align=inside,
        major grid style={densely dotted},
        xticklabel style = {font=\footnotesize, rotate=-90},
        xlabel={\footnotesize \# Units},
]

\addplot+[draw=orange, fill=teal,  fill opacity=0.12] coordinates
	{(12,0)(36,0)(60,0)(84,0.12)(108,0.31)(132,0.65)(168,0.73)(204,1.28)(240,1.71)(344,1.91)(454,2.1)(552,2.11)};
	
\end{axis}
\end{tikzpicture}%
}
\caption{Performance versus complexity trade-off: Q-factor gain of the neural networks over linear equalization as a function of the number of recurrent hidden units in RX 1.  The shaded region is described in the text. 
}
\label{fig: gains-params-RX1}
\end{figure}

The two RX models in Fig. \ref{fig: system-model} were considered for equalization using the proposed CRNN model, as well as MLP, CNN+MLP, bi-RNN, bi-GRU, bi-LSTM, and CNN+bi-LSTM neural structures with similar architectures to \cite{MLP-Yves}, \cite{8386096}, \cite{deligiannidis2021performance}  \cite{liu2021bi}, \cite{deligiannidis2020compensation}, \cite{pedro2021perf&comp}, respectively. The following subsections discuss the performance versus the complexity of these neural network structures in each RX model, and discuss an overall comparison between the two RX types.

\subsection{RX model 1}

In RX model 1, as discussed in Section \ref{sec: principles}, the neural network is placed after CD compensation, RDE-based MIMO equalizer, demultiplexer, and CPE; and in consequence, it operates on 1 SpS signals with the input-output scheme illustrated in Fig. \ref{fig: input-output-scheme}.

The details of the implemented neural networks for RX 1 are mentioned in Table. \ref{tab: model-details-RX1}.
The hyper-parameters in each case, especially the number of hidden units in the recurrent layers, were optimized using K-fold cross-validation, considering the performance-complexity trade-off. Fig. \ref{fig: gains-params-RX1} illustrates the Q-factor gain over linear equalization as a function of the number of hidden units in each cell in the recurrent layers. 
The region of interest is the shaded area between the gain curve and the linear relation, where a linear increase in the number of hidden units yields a higher than linear payoff. 
Note that for bi-directional layers, the presented numbers in the figure and table are the sum of hidden units for both directions, half for each. 
All models were trained using the Adam algorithm in Tensorflow 2.6 with
mean square error (MSE) loss function, $2^{18}$ normalized input output training vectors (with the structure discussed in Section \ref{sec: rx-models-dscrptn}), batch size of 16, learning rate of $5 \times 10^{-4}$, and the decay rates $\beta_1^{Adam}= 8.5 \times 10^{-1}$, $\beta_2^{Adam}=9.99 \times 10^{-1}$, on a Linux Fedora release 35 system with 96 CPUs AMD EPYC 7F72 24-Core Processor, 2 threads per core, and 259 GB RAM. The number of epochs was set to 120. The value of loss on the validation set was calculated at each epoch, and the model with the lowest validation error over epochs was selected.

\begin{figure}[t]
    \centering
    \begin{tikzpicture}
\begin{semilogyaxis}[
    title={},
    ymode=log,
    xlabel={Launch Power $P$ [dBm]},
    ylabel={BER},
    table/col sep=comma,
    xmin=-5, xmax=11,
    xtick={-5,-1,...,11},
    tick label style={font=\large},
    legend pos=outer north east,
    grid=both,
    grid style=dashed,
            legend style={
            font=\small,
            at={(0.5,-0.25)},
            anchor=north,
            legend columns=2,
            /tikz/every even column/.append style={column sep=0.5cm},
            fill opacity=0.7, text opacity = 1
        },
    width=8cm,
     cycle list name=color
]

\addplot+[
    mark=pentagon*,
    densely dashed
    ]
    table[x=Power, y=BER]{linear_Qfactor.csv};
    \addlegendentry{Linear equalization}

\addplot[
    color=teal,
    mark=pentagon*,
    densely dashed
    ]
    table[x=Power, y=BER]{dbp2_Qfactor.csv};
    \addlegendentry{DBP-2StPS}
    
\addplot+[
    mark=oplus*,
    solid
    ]
    table[x=Power, y=BER]{mlp_Qfactor-cpe.csv};
     \addlegendentry{MLP \cite{MLP-Yves}}

\addplot+[
    mark=triangle*,
    solid
    ]
    table[x=Power, y=BER]{cnn_mlp_Qfactor-cpe.csv};
    \addlegendentry{CNN+MLP \cite{8386096}}

\addplot[
    color=lime!50!black,
    mark=halfsquare*,
    solid
    ]
    table[x=Power, y=BER]{bi_rnn_Qfactor-cpe.csv};
    \addlegendentry{bi-RNN \cite{deligiannidis2021performance}}

\addplot+[
    mark=diamond*,
    solid
    ]
    table[x=Power, y=BER]{bi_gru_Qfactor-cpe.csv};
     \addlegendentry{bi-GRU \cite{liu2021bi}}
     
\addplot+[
    mark=triangle*,
    solid
    ]
    table[x=Power, y=BER]{bi_lstm_Qfactor-cpe.csv};
     \addlegendentry{bi-LSTM \cite{deligiannidis2020compensation}}     
    
\addplot[
    color=yellow!50!black,,
    mark=otimes*,
    solid
    ]
    table[x=Power, y=BER]{pedro_Qfactor-cpe.csv};
    \addlegendentry{CNN(no-stride)+bi-LSTM \cite{pedro2021perf&comp}}

\addplot[
    color=green!50!black,
    mark=cube*,
    solid
    ]
    table[x=Power, y=BER]{cnn_rnn_Qfactor-cpe.csv};
     \addlegendentry{\textbf{Proposed CRNN}}
    
\addplot[
    color=red,
    mark=pentagon*,
    densely dashed
    ]
    table[x=Power, y=BER]{dbp80_Qfactor.csv};
    \addlegendentry{DBP-80StPS}     
    
\end{semilogyaxis}
\end{tikzpicture}
    \caption{BER of the neural networks for RX 1 in the test mode as a function of the total launch power.}
    \label{fig: ber-plot-RX1}
\end{figure}
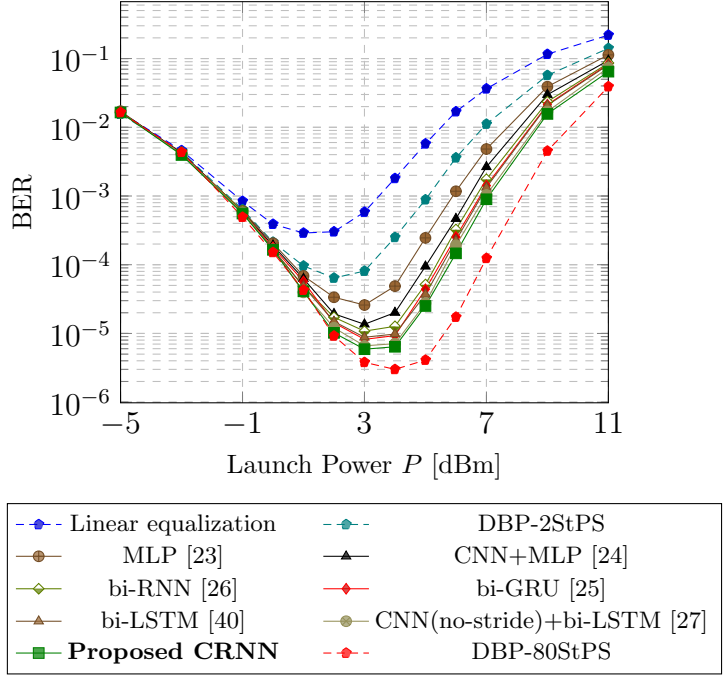

\begin{figure}[t]
    \centering
    \begin{tikzpicture}
\begin{axis}[
    title={},
    xlabel={Launch Power $P$ [dBm]},
    ylabel={Q-factor [dB]},
    table/col sep=comma,
    xmin=-3, xmax=7,
    ymin=6, ymax=14,
    ytick={2,3,...,16},
    xtick={-5,-3,...,11},
    tick label style={font=\large},
    legend pos=outer north east,
    grid=both,
    grid style=dashed,
            legend style={
            font=\small,
            at={(0.5,-0.25)},
            anchor=north,
            legend columns=2,
            /tikz/every even column/.append style={column sep=0.5cm},
            fill opacity=0.7, text opacity = 1
        },
    width=8cm,
     cycle list name=myexotic
]

\addplot[
    color=olive!50!black,
    mark=pentagon*,
    densely dashed
    ]
    table[x=Power, y=Qfactor]{linear_Qfactor.csv};
    \addlegendentry{Linear equalization}

\addplot[
    color=cyan!50!black,
    mark=pentagon*,
    densely dashed
    ]
    table[x=Power, y=Qfactor]{dbp2_Qfactor.csv};
    \addlegendentry{DBP-2StPS}
    
\addplot[
    color=cyan!50!black,
    mark=oplus*,
    solid
    ]
    table[x=Power, y=Qfactor]{mlp_Qfactor-cpe.csv};
     \addlegendentry{MLP \cite{MLP-Yves}}

\addplot+[
    mark=triangle*,
    solid
    ]
    table[x=Power, y=Qfactor]{cnn_mlp_Qfactor-cpe.csv};
    \addlegendentry{CNN+MLP \cite{8386096}}

\addplot[
    color=green!40!black,
    mark=halfsquare*,
    solid
    ]
    table[x=Power, y=Qfactor]{bi_rnn_Qfactor-cpe.csv};
    \addlegendentry{bi-RNN \cite{deligiannidis2021performance}}

\addplot+[
    mark=diamond*,
    solid
    ]
    table[x=Power, y=Qfactor]{bi_gru_Qfactor-cpe.csv};
     \addlegendentry{bi-GRU \cite{liu2021bi}}
     
\addplot[
    color=blue!50!black,
    mark=triangle*,
    solid
    ]
    table[x=Power, y=Qfactor]{bi_lstm_Qfactor-cpe.csv};
     \addlegendentry{bi-LSTM \cite{deligiannidis2020compensation}}     
    
\addplot[
    color=yellow!50!black,,
    mark=otimes*,
    solid
    ]
    table[x=Power, y=Qfactor]{pedro_Qfactor-cpe.csv};
    \addlegendentry{CNN(no-stride)+bi-LSTM \cite{pedro2021perf&comp}}

\addplot[
    color=orange!70!black,
    mark=cube*,
    solid
    ]
    table[x=Power, y=Qfactor]{cnn_rnn_Qfactor-cpe.csv};
     \addlegendentry{\textbf{Proposed CRNN}}
    
\addplot[
    color=brown!50!black,
    mark=pentagon*,
    densely dashed
    ]
    table[x=Power, y=Qfactor]{dbp80_Qfactor.csv};
    \addlegendentry{DBP-80StPS}     

    \addplot[black, no marks, densely dashed]
    coordinates {(-5,6.7) (4,14.85)} node[above, pos = 0.55, rotate=41.75] {linear transmission \(\gamma=0\)};
    
\end{axis}
\end{tikzpicture}
    \caption{Q-factor of the neural networks for RX 1 in the test mode as a function of the total launch power.}
    \label{fig: qfactor-plots-RX1}
\end{figure}
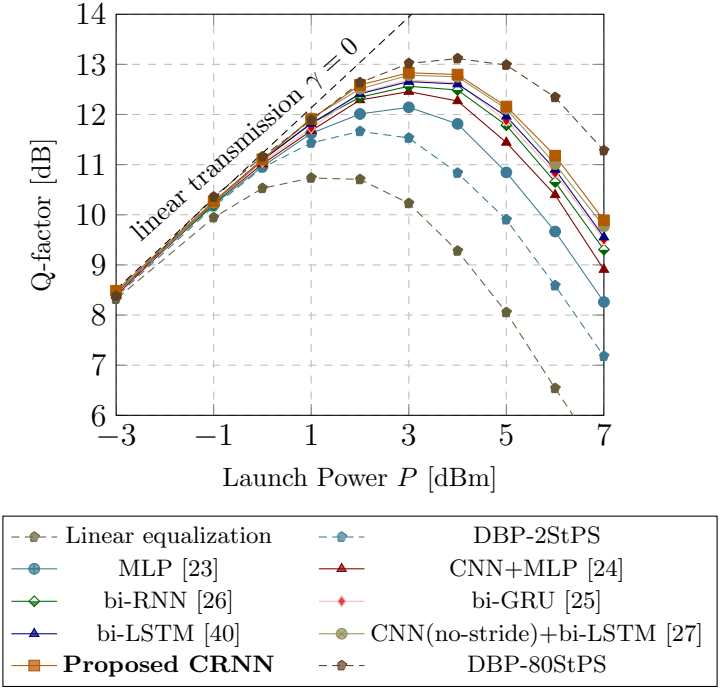

The resulting Q-factor and BER performance plots of the implemented approaches, using $2^{24}$ unseen online-generated bits, are demonstrated in Fig. \ref{fig: ber-plot-RX1}, and Fig. \ref{fig: qfactor-plots-RX1}, respectively. According to these figures, the recurrent-based equalizers provide slightly superior performance over MLP and CNN+MLP models with $0.42-0.69$ dB and $0.11-0.38$ dB Q-factor gain, respectively, at the optimal launch power, and in general $1.83-2.1$ dB gain over linear equalization at optimal launch power. As discussed, it is analyzed that the reason for the lower performance of MLP and CNN+MLP models is because of being susceptible to over-fitting due to the immoderate number of parameters associated with fully-connected hidden layers. Nonetheless, all the neural models achieve relatively close performance to each other with $<0.8$ dB Q-factor difference. It is analyzed that this is in light of the fact that the number of parameters (weights) in the models are relatively limited ($<10^{7}$) because of the limited size of the input. This results in mitigation of the over-fitting risk via an effective training strategy and escalating the likelihood of converging to the same low bias level via a suitable network complexity, which on the contrary, was investigated to be totally different for different models. 

\pgfplotscreateplotcyclelist{myexotic}{%
teal!50!black,every mark/.append style={fill=teal!80!black},mark=*\\%
orange!50!black,every mark/.append style={fill=orange!80!black},mark=square*\\%
cyan!70!black,every mark/.append style={fill=cyan!80!black},mark=otimes*\\%
red!50!black,mark=star\\%
lime!90!black,every mark/.append style={fill=lime},mark=diamond*\\%
red!30!white,densely dashed,every mark/.append style={solid,fill=red!80!black},mark=*\\%
yellow!60!black,densely dashed,
every mark/.append style={solid,fill=yellow!80!black},mark=square*\\%
black,every mark/.append style={solid,fill=gray},mark=otimes*\\%
blue,densely dashed,mark=star,every mark/.append style=solid\\%
red,densely dashed,every mark/.append style={solid,fill=red!80!black},mark=diamond*\\%
}

\begin{table}[!ht]
\caption{Neural networks' number of parameters and FLOPs per symbol, in RX 1. }
\vspace{-0.35cm}
\small
\center
\begin{tabular}{@{}ccc@{}}
\toprule
            &           &                           \\
\multirow{-2}{*}{\textbf{Model}} & \multirow{-2}{*}{\textbf{\#Params}} & \multirow{-2}{*}{\textbf{\#FLOPs/symbol}}        \\ \midrule \midrule
MLP  \cite{MLP-Yves}       & 5,316,100 & ${\sim}26.90 \times 10^5$ \\ \midrule
CNN+MLP   \cite{8386096}   & 3,940,966 & ${\sim}21.79 \times 10^5$ \\ \midrule
bi-LSTM \cite{deligiannidis2020compensation}    & 154,948   & ${\sim}54.61 \times 10^5$ \\ \midrule
bi-GRU \cite{liu2021bi}     & 144,292   & ${\sim}39.81 \times 10^5$ \\ \midrule
bi-RNN \cite{deligiannidis2021performance}     & 275,044   & ${\sim}33.40 \times 10^5$ \\ \midrule
CNN+bi-LSTM [27] & 139.066   & ${\sim}47.39 \times 10^5$ \\ \midrule
\textbf{Proposed CRNN}        & 324,418                             & {\color[HTML]{9A0000} ${\sim} \mathbf{\normalsize 16.47} \times 10^5$ }\\ \bottomrule
\end{tabular}
\label{tab: RX1-nns-comp}
\end{table}

\begin{table}[t]
\caption{Details of the hidden layers of the implemented neural networks for RX 2. The output layer for all the models is a linear fully-connected layer with 4 units. 
}
\centering
\footnotesize
\begin{tabular}{@{}cccc@{}}
\toprule
\textbf{Model}           & \textbf{Layer} & \textbf{Type} & \textbf{Details}                                              \\ \midrule
\multirow{5}{*}{MLP \cite{MLP-Yves}}     & layer 1        & Dropout       & dropout rate: 0.5                                             \\ \cmidrule(l){2-4} 
                         & layer 2        & FC            & \#units: 1536,  activ.: tanh                                  \\ \cmidrule(l){2-4} 
                         & layer 3,5      & FC            & \#units: 1152,  activ.: tanh                                  \\ \cmidrule(l){2-4} 
                         & layer 4        & Dropout       & dropout rate: 0.4                                             \\ \cmidrule(l){2-4} 
                         & layer 6        & FC            & \#units: 768,  activ.: tanh                                   \\ \midrule
\multirow{5}{*}{CNN+MLP \cite{8386096}}                                                     & layer 1 & Conv-1D & $\textnormal{L}_{ker}$:  99, strd: 1, \#ch.: 3, activ.: relu        \\ \cmidrule(l){2-4} 
                         & layer 2        & Conv-1D       & $\textnormal{L}_{ker}$:  99, strd: 3, \#ch.: 7, activ.: relu  \\ \cmidrule(l){2-4} 
                         & layer 3        & Conv-1D       & $\textnormal{L}_{ker}$:  51, strd: 3, \#ch.: 20, activ.: relu \\ \cmidrule(l){2-4} 
                         & layer 5        & Dropout       & dropout rate: 0.3                                             \\ \cmidrule(l){2-4} 
                         & layer 4,6      & FC            & \#units: 1152, activ.: tanh                                   \\ \midrule
\multirow{2}{*}{bi-LSTM \cite{deligiannidis2020compensation}} & layer 1        & bi-LSTM       & \#units: 158; activ.: tanh, sigmoid                           \\ \cmidrule(l){2-4} 
                         & layer 2        & Flattening    & data format: channel last                                     \\ \midrule
\multirow{2}{*}{bi-GRU \cite{liu2021bi}}  & layer 1        & bi-GRU        & \#units: 158; activ.: tanh, sigmoid                           \\ \cmidrule(l){2-4} 
                         & layer 2        & Flattening    & data format: channel last                                     \\ \midrule
\multirow{2}{*}{bi-RNN \cite{deligiannidis2021performance}}  & layer 1        & bi-RNN        & \#units: 270; activ.: tanh                                    \\ \cmidrule(l){2-4} 
                         & layer 2        & Flattening    & data format: channel last                                     \\ \midrule
\multirow{3}{*}{CNN+bi-LSTM \cite{pedro2021perf&comp}}                                                 & layer 1 & Conv-1D & $\textnormal{L}_{ker}$:  49, strd: 1, \#ch.: 2,  activ.: leaky relu \\ \cmidrule(l){2-4} 
                         & layer 2        & bi-LSTM       & \#units: 136,  activ.: tanh, sigmoid                          \\ \cmidrule(l){2-4} 
                         & layer 3        & Flattening    & data format: channel last                                     \\ \midrule
\multirow{4}{*}{\begin{tabular}[c]{@{}c@{}} \textbf{Proposed CRNN} \end{tabular}} & layer 1 & Conv-1D & $\textnormal{L}_{ker}$: 99, strd: 1, \#ch.: 4, activ: relu          \\ \cmidrule(l){2-4} 
                         & layer 2        & Conv-1D       & $\textnormal{L}_{ker}$: 11, strd: 11, \#ch.: 30, activ: relu  \\ \cmidrule(l){2-4} 
                         & layer 3        & Conv 1-D      & $\textnormal{L}_{ker}$:  9, strd: 9, \#ch.: 125, activ: relu  \\ \cmidrule(l){2-4} 
                         & layer 4        & uni-RNN       & \#units: 494, activ.: tanh                                    \\ \bottomrule
\end{tabular}
\label{tab: model-details-RX2}
\end{table}

Table. \ref{tab: RX1-nns-comp} presents the complexity of the implemented neural network models as the number of FLOPs they incur in the inference mode per output symbol. These values are obtained analytically according to the formulas discussed in Section \ref{sec: recurrent-networks} for the recurrent layers, the following formula for the fully connected layers 
\begin{equation}
    \textnormal{FLOPs}^{\{\textnormal{fully-connected}\}} = n_in_h + \eta n_h + n_h, 
\end{equation}
where $n_i$ is the number of input features and $n_h$ is the number of hidden units, and the following formula for the convolution layers
\begin{equation}
    \textnormal{FLOPs}^{\{\textnormal{conv}\}} = n_{ker} \times (2 \times e_{ker}-1) \times L_{out} + \eta e_{out},
\label{eq: FLOPS-conv}
\end{equation}
where $n_{ker}$ is the number of kernels, $e_{ker}$ is the kernel shape, and $e_{out}$ is the shape of output feature-map. $L_{out}$, which is the length of output feature-map, is obtained by 
\begin{equation}
    L_{out} = \Bigl\lfloor \frac{L_{in}+2 \times pad - (dil \times L_{ker}-1)-1 }{strd} +1 \Bigr\rfloor,
\label{eq: FLOPS-cnn}
\end{equation}
where $pad, dil$ and $strd$, respectively, signify padding, dilation, and stride. $L_{ker}$ is the kernel length. The values are also checked with their proportionality to the number of CPU cycles in the training mode obtained by the Linux kernel performance monitoring tool \emph{perf} library. A full report of the CPU performance counters and trace-points recorded for each neural network is available at \cite{complexity-report-kaggle}.
As Table. \ref{tab: RX1-nns-comp} demonstrates, although the proposed CRNN achieves a comparable performance to the bi-RNN based models, it has $>50\%$ lower complexity than these models, namely ${\sim} 50.6 \%$, ${\sim} 58.6 \%$, ${\sim} 69.8 \%$, and ${\sim} 65.2 \%$ fewer number of FLOPs than bi-RNN, bi-GRU, bi-LSTM, and CNN+bi-LSTM models; which is thanks to the efficiency of the model in minimizing the overlap between short-range and long-range dependencies extraction, having quite limited number of time-steps for the recurrent layer allowing for leveraging uni-directional many-to-one recurrent layer instead of bi-directional, and using vanilla RNN in place of gated recurrent cells.

\subsection{RX model 2}

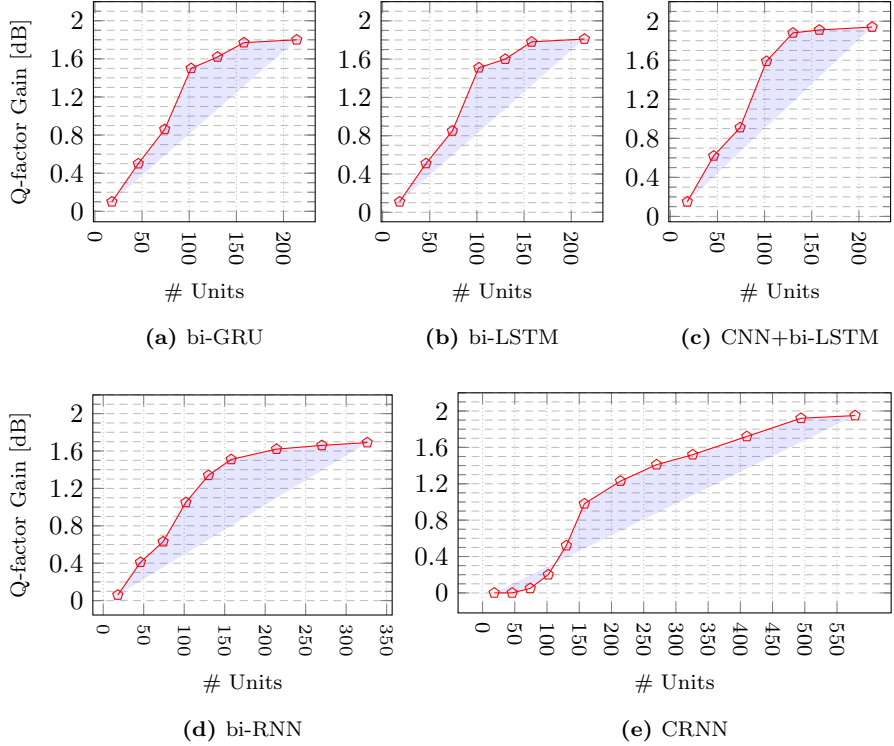
\begin{figure}[t]
\centering

\subfloat[bi-GRU]{%
\begin{tikzpicture}[trim axis left,trim axis right]
\begin{axis}[
        height=4.5cm, width=4.5cm,
        yminorgrids, tick align=inside,
        minor grid style={densely dashed},
        ymax=2.2,
        ytick={0,0.4,0.8,1.2,1.6,2.0},
        minor ytick={0,0.1,0.2,0.3,0.4,0.5,0.6,0.7,0.8,0.9,1.0,1.1,1.2,1.3,1.4,1.5,1.6,1.7,1.8,1.9,2.0,2.1,2.2},
        xtick={0,50,...,350},
                xmajorgrids, tick align=inside,
        major grid style={densely dotted},
        xticklabel style = {font=\footnotesize, rotate=-90},
                ylabel={\footnotesize Q-factor Gain [dB]},
        xlabel={\footnotesize \# Units},
]
\addplot+[mark=pentagon*, draw=red, fill=blue,  fill opacity=0.1] coordinates
	{(18,0.1)(46,0.50)(74,0.86)(102,1.50)(130,1.62)(158,1.77)(214,1.80)};
\end{axis}
\end{tikzpicture}%
}
\hspace{0.75cm}
\subfloat[bi-LSTM]{%
\begin{tikzpicture}[trim axis left,trim axis right]
\begin{axis}[
        height=4.5cm, width=4.5cm,
        yminorgrids, tick align=inside,
        minor grid style={densely dashed},
        ymax=2.2,
        ytick={0,0.4,0.8,1.2,1.6,2.0},
        minor ytick={0,0.1,0.2,0.3,0.4,0.5,0.6,0.7,0.8,0.9,1.0,1.1,1.2,1.3,1.4,1.5,1.6,1.7,1.8,1.9,2.0,2.1,2.2},
        xtick={0,50,...,350},
                xmajorgrids, tick align=inside,
        major grid style={densely dotted},
        xticklabel style = {font=\footnotesize, rotate=-90},
        xlabel={\footnotesize \# Units},
]

\addplot+[mark=pentagon*, draw=red, fill=blue,  fill opacity=0.1] coordinates
	{(18,0.11)(46,0.51)(74,0.85)(102,1.51)(130,1.60)(158,1.78)(214,1.81)};
	
\end{axis}
\end{tikzpicture}%
}
\hspace{0.75cm}
\subfloat[CNN+bi-LSTM]{%
\begin{tikzpicture}[trim axis left,trim axis right]
\begin{axis}[
        height=4.5cm, width=4.5cm,
        yminorgrids, tick align=inside,
        minor grid style={densely dashed},
        ymax=2.2,
        ytick={0,0.4,0.8,1.2,1.6,2.0},
        minor ytick={0,0.1,0.2,0.3,0.4,0.5,0.6,0.7,0.8,0.9,1.0,1.1,1.2,1.3,1.4,1.5,1.6,1.7,1.8,1.9,2.0,2.1,2.2},
        xtick={0,50,...,350},
                xmajorgrids, tick align=inside,
        major grid style={densely dotted},
        xticklabel style = {font=\footnotesize, rotate=-90},
        xlabel={\footnotesize \# Units},
]

\addplot+[mark=pentagon*, draw=red, fill=blue,  fill opacity=0.1] coordinates
	{(18,0.15)(46,0.62)(74,0.91)(102,1.59)(130,1.88)(158,1.91)(214,1.94)};
	
\end{axis}
\end{tikzpicture}%
}

\vspace{0.5cm}

\subfloat[bi-RNN]{%
\begin{tikzpicture}[trim axis left,trim axis right]
\begin{axis}[
        height=4.5cm, width=5.52cm,
        yminorgrids, tick align=inside,
        minor grid style={densely dashed},
        ymax=2.2,
        ytick={0,0.4,0.8,1.2,1.6,2.0},
        minor ytick={0,0.1,0.2,0.3,0.4,0.5,0.6,0.7,0.8,0.9,1.0,1.1,1.2,1.3,1.4,1.5,1.6,1.7,1.8,1.9,2.0,2.1,2.2},
        xtick={0,50,...,350},
        xmajorgrids, tick align=inside,
        major grid style={densely dotted},
        xticklabel style = {font=\scriptsize, rotate=-90},
        ylabel={\footnotesize Q-factor Gain [dB]},
        ylabel near ticks,
        xlabel={\footnotesize \# Units},
]
\addplot+[mark=pentagon*, draw=red, fill=blue,  fill opacity=0.1] coordinates
	{(18,0.06)(46,0.41)(74,0.63)(102,1.05)(130,1.34)(158,1.51)(214,1.62)(270,1.66)(326,1.69)};
\end{axis}
\end{tikzpicture}%
}
\hspace{0.75cm}
\subfloat[CRNN]{%
\begin{tikzpicture}[trim axis left,trim axis right]
\begin{axis}[
        height=4.5cm, width=7.28cm,
        yminorgrids, tick align=inside,
        minor grid style={densely dashed},
        ymax=2.2,
        ytick={0,0.4,0.8,1.2,1.6,2.0},
        minor ytick={0,0.1,0.2,0.3,0.4,0.5,0.6,0.7,0.8,0.9,1.0,1.1,1.2,1.3,1.4,1.5,1.6,1.7,1.8,1.9,2.0,2.1,2.2},
        xtick={0,50,...,550},
        xmajorgrids, tick align=inside,
        major grid style={densely dotted},
        xticklabel style = {font=\footnotesize, rotate=-90},
        xlabel={\footnotesize \# Units},
]

\addplot+[mark=pentagon, draw=red, fill=blue,  fill opacity=0.1] coordinates
	{(18,0)(46,0)(74,0.05)(102,0.20)(130,0.52)(158,0.98)(214,1.23)(270,1.41)(326,1.52)(410,1.72)(494,1.92)(578,1.95)};
	
\end{axis}
\end{tikzpicture}%
}
\caption{Performance versus complexity trade-off: Q-factor gain of the neural networks over linear equalization as a function of the number of recurrent hidden units in RX 2. The shaded region is described in the text.}
\label{fig: gains-params-RX2}
\end{figure}

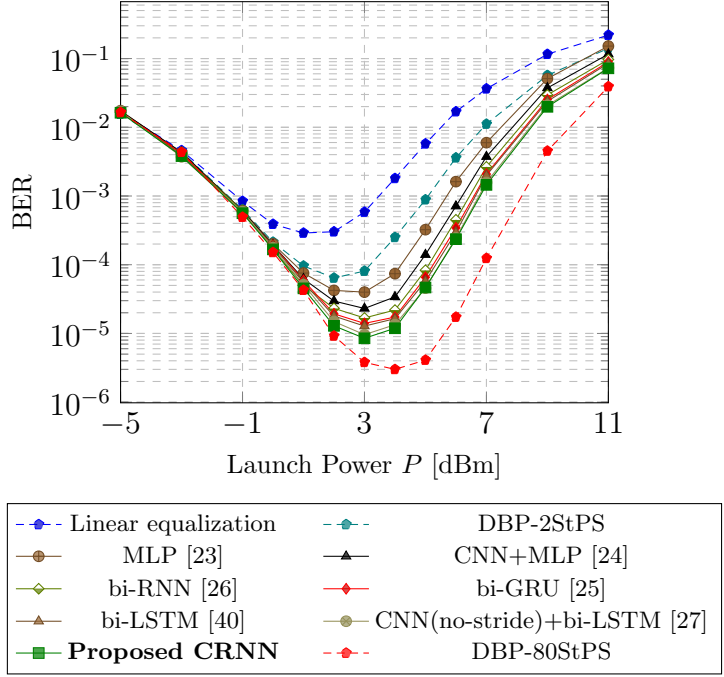
\begin{figure}[t]
    \centering
    \begin{tikzpicture}
\begin{semilogyaxis}[
    title={},
    ymode=log,
    xlabel={Launch Power $P$ [dBm]},
    ylabel={BER},
    table/col sep=comma,
    xmin=-5, xmax=11,
    xtick={-5,-1,...,11},
    tick label style={font=\large},
    legend pos=outer north east,
    grid=both,
    grid style=dashed,
    legend style={
            font=\small,
            at={(0.5,-0.25)},
            anchor=north,
            legend columns=2,
            /tikz/every even column/.append style={column sep=0.5cm},
            fill opacity=0.7, text opacity = 1
    },
    width=8cm,
     cycle list name=color
]

\addplot+[
    mark=pentagon*,
    densely dashed
    ]
    table[x=Power, y=BER]{linear_Qfactor.csv};
    \addlegendentry{Linear equalization}

\addplot[
    color=teal,
    mark=pentagon*,
    densely dashed
    ]
    table[x=Power, y=BER]{dbp2_Qfactor.csv};
    \addlegendentry{DBP-2StPS}
    
\addplot+[
    mark=oplus*,
    solid
    ]
    table[x=Power, y=BER]{mlp_Qfactor-e2e.csv};
     \addlegendentry{MLP \cite{MLP-Yves}}

\addplot+[
    mark=triangle*,
    solid
    ]
    table[x=Power, y=BER]{cnn_mlp_Qfactor-e2e.csv};
    \addlegendentry{CNN+MLP \cite{8386096}}

\addplot[
    color=lime!50!black,
    mark=halfsquare*,
    solid
    ]
    table[x=Power, y=BER]{bi_rnn_Qfactor-e2e.csv};
    \addlegendentry{bi-RNN \cite{deligiannidis2021performance}}

    
\addplot+[
    mark=diamond*,
    solid
    ]
    table[x=Power, y=BER]{bi_gru_Qfactor-e2e.csv};
     \addlegendentry{bi-GRU \cite{liu2021bi}}
     
\addplot+[
    mark=triangle*,
    solid
    ]
    table[x=Power, y=BER]{bi_lstm_Qfactor-e2e.csv};
     \addlegendentry{bi-LSTM \cite{deligiannidis2020compensation}}     
    
\addplot[
    color=yellow!50!black,,
    mark=otimes*,
    solid
    ]
    table[x=Power, y=BER]{pedro_Qfactor-e2e.csv};
    \addlegendentry{CNN(no-stride)+bi-LSTM \cite{pedro2021perf&comp}}

\addplot[
    color=green!50!black,
    mark=cube*,
    solid
    ]
    table[x=Power, y=BER]{cnn_rnn_Qfactor-e2e.csv};
     \addlegendentry{\textbf{Proposed CRNN}}
    
\addplot[
    color=red,
    mark=pentagon*,
    densely dashed
    ]
    table[x=Power, y=BER]{dbp80_Qfactor.csv};
    \addlegendentry{DBP-80StPS}     
    
\end{semilogyaxis}
\end{tikzpicture}
    \caption{BER of the neural networks for RX 2 in the test mode as a function of the total launch power.}
    \label{fig: ber-plot-RX2}
\end{figure}

\pgfplotscreateplotcyclelist{myexotic}{%
teal!50!black,every mark/.append style={fill=teal!80!black},mark=*\\%
orange!50!black,every mark/.append style={fill=orange!80!black},mark=square*\\%
cyan!70!black,every mark/.append style={fill=cyan!80!black},mark=otimes*\\%
red!50!black,mark=star\\%
lime!90!black,every mark/.append style={fill=lime},mark=diamond*\\%
red!30!white,densely dashed,every mark/.append style={solid,fill=red!80!black},mark=*\\%
yellow!60!black,densely dashed,
every mark/.append style={solid,fill=yellow!80!black},mark=square*\\%
black,every mark/.append style={solid,fill=gray},mark=otimes*\\%
blue,densely dashed,mark=star,every mark/.append style=solid\\%
red,densely dashed,every mark/.append style={solid,fill=red!80!black},mark=diamond*\\%
}

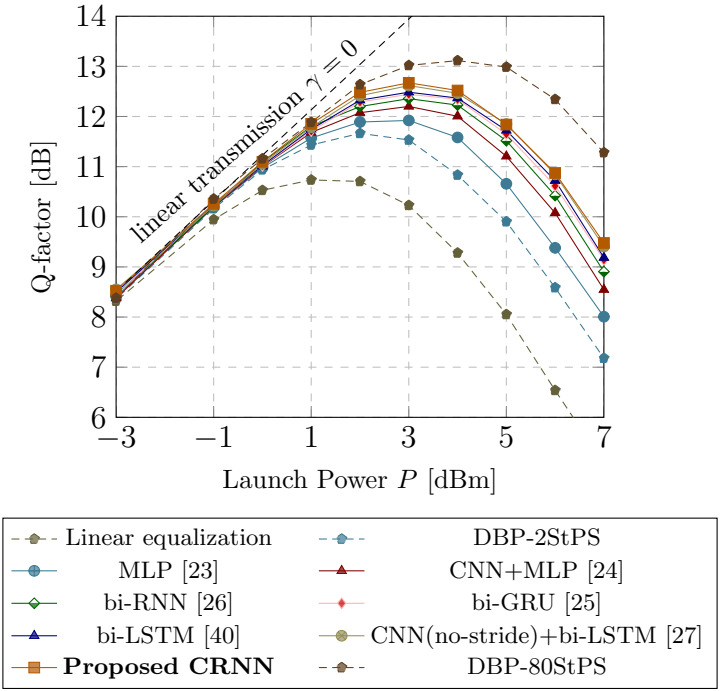
\begin{figure}[t]
    \centering
    \begin{tikzpicture}
\begin{axis}[
    title={},
    xlabel={Launch Power $P$ [dBm]},
    ylabel={Q-factor [dB]},
    table/col sep=comma,
    xmin=-3, xmax=7,
    ymin=6, ymax=14,
    ytick={2,3,...,16},
    xtick={-5,-3,...,11},
    tick label style={font=\large},
    legend pos=outer north east,
    grid=both,
    grid style=dashed,
    legend style={
            font=\small,
            at={(0.5,-0.25)},
            anchor=north,
            legend columns=2,
            /tikz/every even column/.append style={column sep=0.5cm},
            fill opacity=0.7, text opacity = 1
    },
    width=8cm,
     cycle list name=myexotic
]

\addplot[
    color=olive!50!black,
    mark=pentagon*,
    densely dashed
    ]
    table[x=Power, y=Qfactor]{linear_Qfactor.csv};
    \addlegendentry{Linear equalization}

\addplot[
    color=cyan!50!black,
    mark=pentagon*,
    densely dashed
    ]
    table[x=Power, y=Qfactor]{dbp2_Qfactor.csv};
    \addlegendentry{DBP-2StPS}
    
\addplot[
    color=cyan!50!black,
    mark=oplus*,
    solid
    ]
    table[x=Power, y=Qfactor]{mlp_Qfactor-e2e.csv};
     \addlegendentry{MLP \cite{MLP-Yves}}

\addplot+[
    mark=triangle*,
    solid
    ]
    table[x=Power, y=Qfactor]{cnn_mlp_Qfactor-e2e.csv};
    \addlegendentry{CNN+MLP \cite{8386096}}

\addplot[
    color=green!40!black,
    mark=halfsquare*,
    solid
    ]
    table[x=Power, y=Qfactor]{bi_rnn_Qfactor-e2e.csv};
    \addlegendentry{bi-RNN \cite{deligiannidis2021performance}}

\addplot+[
    mark=diamond*,
    solid
    ]
    table[x=Power, y=Qfactor]{bi_gru_Qfactor-e2e.csv};
     \addlegendentry{bi-GRU \cite{liu2021bi}}
     
\addplot[
    color=blue!50!black,
    mark=triangle*,
    solid
    ]
    table[x=Power, y=Qfactor]{bi_lstm_Qfactor-e2e.csv};
     \addlegendentry{bi-LSTM \cite{deligiannidis2020compensation}}     
    
\addplot[
    color=yellow!50!black,,
    mark=otimes*,
    solid
    ]
    table[x=Power, y=Qfactor]{pedro_Qfactor-e2e.csv};
    \addlegendentry{CNN(no-stride)+bi-LSTM \cite{pedro2021perf&comp}}

\addplot[
    color=orange!70!black,
    mark=cube*,
    solid
    ]
    table[x=Power, y=Qfactor]{cnn_rnn_Qfactor-e2e.csv};
     \addlegendentry{\textbf{Proposed CRNN}}
    
\addplot[
    color=brown!50!black,
    mark=pentagon*,
    densely dashed
    ]
    table[x=Power, y=Qfactor]{dbp80_Qfactor.csv};
    \addlegendentry{DBP-80StPS}     

    \addplot[black, no marks, densely dashed]
    coordinates {(-5,6.7) (4,14.85)} node[above, pos = 0.55, rotate=41.75] {linear transmission \(\gamma=0\)};
    
\end{axis}
\end{tikzpicture}
    \caption{Q-factor of the neural networks for RX 2 in the test mode as a function of the total launch power.}
    \label{fig: qfactor-plots-RX2}
\end{figure}

As discussed, the neural network in RX 2 is responsible for joint nonlinearity-PMD compensation upon receiving the sampled waveform after CD compensation with 2 SpS. As the signal is subject to the random phase noise effect, a CPE block is required outside the neural network to compensate for the phase noise. Table. \ref{tab: model-details-RX2} presents the details of the implemented neural networks for RX 2. The networks were tuned (Fig. \ref{fig: gains-params-RX2}), trained, and tested using the same scheme and number of samples as for the neural networks in RX 1. On the basis of the discussion in Section \ref{sec: rx-models-dscrptn}, the neural networks were trained using the output of the CD compensation at 2 SpS, and the ground truth equal to the corresponding correct symbols at TX in the standard 16-QAM constellation.

Fig. \ref{fig: ber-plot-RX2} and Fig. \ref{fig: qfactor-plots-RX2} demonstrate the resulting BER and Q-factor performance of the models. As it is noticeable in comparison with the corresponding plots for RX 1, the performance in RX 2 is diminished (averagely ${\sim}0.2$ dB), which is due to the destructive effect of unmitigated random phase noise in the training process. In RX 2, similarly, although the neural networks achieve roughly comparable performance as it is shown, they incur substantially different complexities. Table. \ref{tab: RX2-nns-comp} demonstrates the number of FLOPs incurred by each neural model in RX 2. As this table reports, the proposed CRNN incurs ${\sim}52.1\%$, ${\sim}58.1\%$, ${\sim}69.7\%$, ${\sim}65.4\%$ lower complexity than bi-RNN, bi-GRU, bi-LSTM, and CNN+bi-LSTM models, respectively, which are roughly escalated ratios compared to that calculated for RX 1; in light of the same underlying reasons, plus the dimensionality reduction prior to the recurrent layer.              

\begin{table}[!ht]
\caption{Neural networks' number of parameters and FLOPs per symbol, in RX 2. }
\vspace{-0.35cm}
\small
\center
\begin{tabular}{@{}ccc@{}}
\toprule
            &           &                           \\
\multirow{-2}{*}{\textbf{Model}}                                     & \multirow{-2}{*}{\#Params} & \multirow{-2}{*}{\#FLOPs/symbol}                 \\ \midrule \midrule
MLP \cite{MLP-Yves}         & 6,348,292 & ${\sim}32.06 \times 10^5$ \\ \midrule
CNN+MLP \cite{8386096}    & 2,680,183 & ${\sim}26.87 \times 10^5$ \\ \midrule
bi-LSTM \cite{deligiannidis2020compensation}    & 176,964   & ${\sim}66.53 \times 10^5$ \\ \midrule
bi-GRU \cite{liu2021bi}      & 163,534   & ${\sim}48.68 \times 10^5$ \\ \midrule
bi-RNN \cite{deligiannidis2021performance}      & 324,814   & ${\sim}42.50 \times 10^5$ \\ \midrule
CNN+bi-LSTM [27] & 166,002   & ${\sim}63.99 \times 10^5$ \\ \midrule
\textbf{\begin{tabular}[c]{@{}c@{}}Proposed CRNN\end{tabular}} & 344,281                    & {\color[HTML]{9A0000} ${\sim} \mathbf{\normalsize 20.36} \times 10^5$} \\ \bottomrule
\end{tabular}
\label{tab: RX2-nns-comp}
\end{table}

Note that, as discussed in Section \ref{sec: solution}, in RX 2, the neural networks need to be retrained online frequently due to the fiber birefringence. The neural networks need ${\sim}10^{5}$ SGD iterations to be retrained. By assuming the number of FLOPs for one SGD iteration to be 4-6 times that of one forward propagation step, and the required retraining frequency to be $10^{3}$ retrain/s, an overhead of $8-12 \%$ on top of the complexities discussed in Table. \ref{tab: RX2-nns-comp}, should be considered for the models, in addition to the high-speed memory requirement.    

\subsection{RX 1 versus RX 2}

\begin{figure}[t]
    \centering
    \begin{tikzpicture}
\begin{axis}[
    title={},
    xlabel={Launch Power $P$ [dBm]},
    ylabel={Q-factor [dB]},
    table/col sep=comma,
    xmin=-5, xmax=9,
    ymin=6, ymax=14,
    ytick={2,3,...,16},
    xtick={-5,-3,...,11},
    tick label style={font=\large},
    legend pos=outer north east,
    grid=both,
    grid style=dashed,
    legend style={
            font=\small,
            at={(0.5,-0.25)},
            anchor=north,
            legend columns=2,
            /tikz/every even column/.append style={column sep=0.5cm},
            fill opacity=0.7, text opacity = 1
    },
    width=8cm,
     cycle list name=myexotic
]

\addplot[
    color=olive!50!black,
    mark=pentagon*,
    densely dashed
    ]
    table[x=Power, y=Qfactor]{linear_Qfactor.csv};
    \addlegendentry{Linear equalization}

\addplot[
    color=cyan!50!black,
    mark=pentagon*,
    densely dashed
    ]
    table[x=Power, y=Qfactor]{dbp2_Qfactor.csv};
    \addlegendentry{DBP-2StPS}
    
\addplot+[
    color=olive,
    mark=diamond*,
    solid
    ]
    table[x=Power, y=Qfactor]{cnn+rnn_Qfactor-e2e.csv};
     \addlegendentry{RX 1}

\addplot[
    color=orange!80!black,
    mark=triangle*,
    solid
    ]
    table[x=Power, y=Qfactor]{cnn+rnn_Qfactor-e2e-nophase.csv};
    \addlegendentry{RX 2 - no PN in train}

\addplot[
    color=green!80!black,
    mark=triangle*,
    solid
    ]
    table[x=Power, y=Qfactor]{cnn+rnn_Qfactor-cpe.csv};
    \addlegendentry{RX 2}

\addplot[
    color=brown!50!black,
    mark=pentagon*,
    densely dashed
    ]
    table[x=Power, y=Qfactor]{dbp80_Qfactor.csv};
    \addlegendentry{DBP-80StPS}     

    \addplot[black, no marks, densely dashed]
    coordinates {(-5,6.7) (4,14.85)} node[above, pos = 0.5, rotate=51] {linear transmission \(\gamma=0\)};
    
\end{axis}
\end{tikzpicture}
    \caption{Comparison of the Q-factor performance of RX 1 and RX 2 in the test mode.}
    \label{fig: comp-qfactor-plots}
\end{figure}
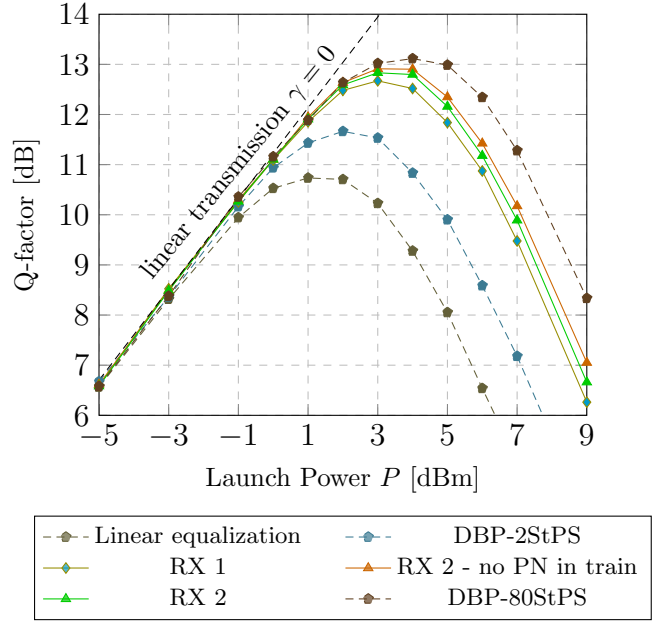

\begin{figure}
 \centering
\begin{subfigure}{0.5\textwidth}
\centering
\includegraphics[width=0.49\textwidth]{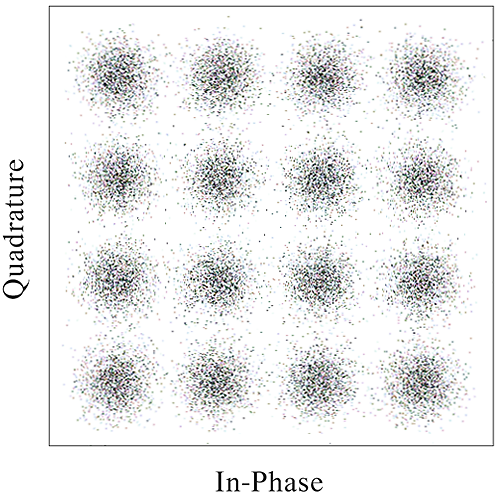}
\caption{At neural network input}
\label{fig: cnstl-linear}
\end{subfigure}
\begin{subfigure}{0.5\textwidth}
\centering
\includegraphics[width=0.48\textwidth]{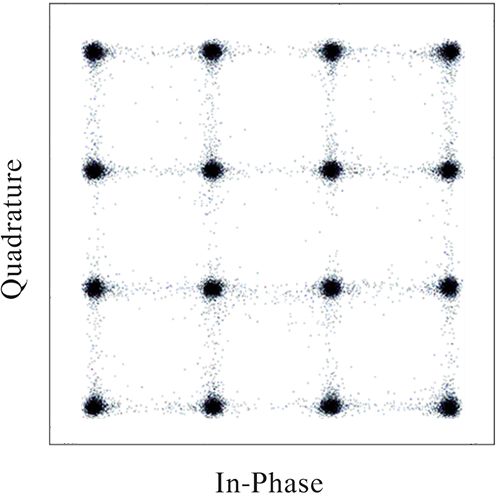}
\caption{At neural network output}
\label{fig: cnstl-rx1-nn}
\end{subfigure}
\caption{Constellations in RX 1.
}

\label{fig: cnstl-RX1}
\end{figure}

\begin{figure}
 \centering
\begin{subfigure}{0.33\textwidth}
\centering
\includegraphics[width=0.75\textwidth]{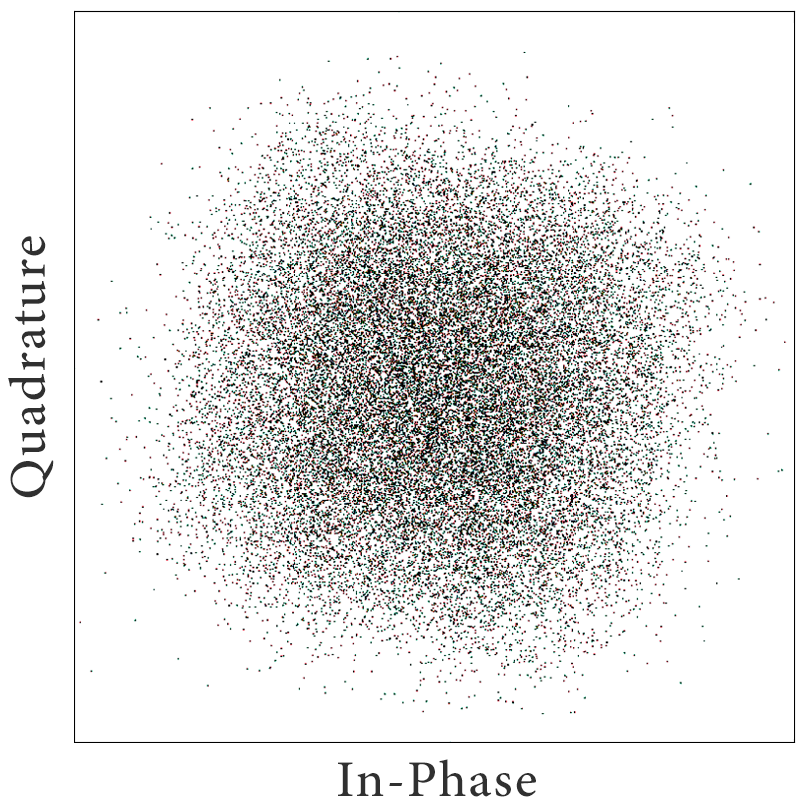}
\caption{At neural network input}
\label{fig: cnstl-rx2-CD}
\end{subfigure} 
\begin{subfigure}{0.33\textwidth}
\centering
\includegraphics[width=0.75\textwidth]{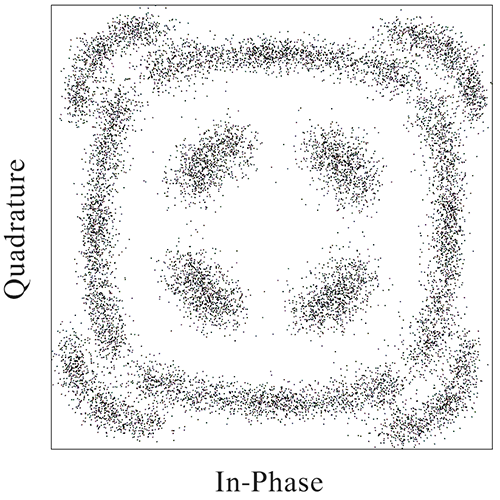}
\caption{At neural network output}
\label{fig: cnstl-rx2-nn}
\end{subfigure}
\begin{subfigure}{0.33\textwidth}
\centering
\includegraphics[width=0.75\textwidth]{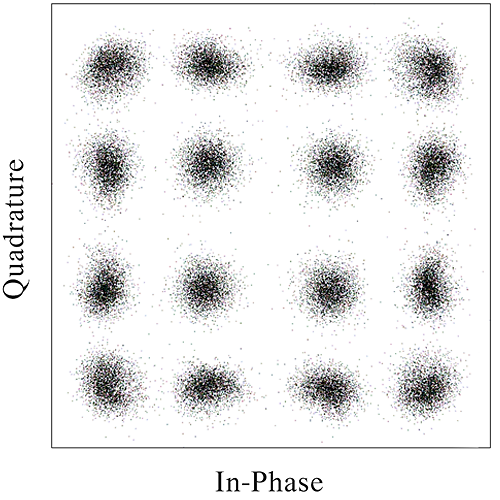}
\caption{After CPE}
\label{fig: cnstl-rx2-cpe}
\end{subfigure}

\caption{Constellations in RX 2.
}
\label{fig: cnstl-RX2}
\end{figure}

The neural networks in RX 1 result in an average 0.2 dB superior Q-factor performance at optimal launch power compared to the neural networks in RX 2, as it is demonstrated in Fig. \ref{fig: comp-qfactor-plots}. 
We also investigated the case where the neural networks are trained using signals not affected by phase noise 
(which is not a realistic assumption), but tested with phase noise.
In this scenario,  RX 2 demonstrates a negligible performance improvement over RX 1 in the test mode. In short, while not having a higher performance than RX 1, the neural networks in RX 2 incur substantially higher complexities than RX 1, in addition to a frequent retraining requirement overhead, which is disproportionate to the number of FLOPs they pull out by eliminating the MIMO equalizer in the DSP chain. In view of this matter, it is believed that RX 1 is a better model to be adopted for deployment in fiber-optic transmission systems. 

We also note that RX 1 and RX 2 result in different constellation diagrams. 
As Fig.~\ref{fig: cnstl-RX1} shows, the constellation in RX 1 forms a square grid or ``jail window'' shape. This effect, which is commonly observed in other work \cite{9125673,6975096,8301433,8868069,kotlyar2021convolutional}, occurs when the neural network equalizer is trained based on regression using the MSE loss function.
This, however, differs from the constellation obtained in RX 2. In RX 2, because of the random phase noise effect, persistent patterns do not exist in the input to be learned for detection. However, some forms of detection in the context of phase noise affected constellation happens, which results in non-uniform noise around the transmitted symbols. Fig. \ref{fig: cnstl-RX2} illustrates a sample constellation after the neural network and after the CPE block in RX 2.

\section{Conclusions}
\label{sec: conclusions}

We proposed a hybrid neural network equalizer comprising a CNN encoder and a unidirectional many-to-one vanilla RNN working in tandem to mitigate nonlinear channel impairments in long-haul fiber-optic communications. 
We showed that the suggested CNN encoder structure not only minimizes the overlap between short-term and long-term dependencies extraction but also, by taking data into a latent space through multi-channeling and striding properties, it reduces the number of significant features so that the subsequent layer can be implemented efficiently in the form of a unidirectional vanilla recurrent layer in lieu of a bidirectional gated recurrent layer.
We demonstrated that for 64 GBd DP-16-QAM optical transmission over $14 \times 80$ km SSMF, the proposed CRNN-based approach achieves the performance of the state-of-the-art bidirectional recurrent-based methods with $>50\%$ lower computational complexity compared to them.

\end{document}